\documentclass[article,nojss,shortnames]{jss}

\usepackage{thumbpdf}
\usepackage{amsfonts,amstext,amsmath,amssymb,amsthm, mathtools}
\usepackage{accents}
\usepackage{rotating}
\usepackage{verbatim}
\usepackage{booktabs}
\usepackage{makecell}
\usepackage{caption}
\usepackage{multirow}
\usepackage{relsize}
\usepackage{diagbox}
\usepackage{numprint}
\usepackage{dcolumn}
\usepackage{pifont}
\usepackage{alltt}
\usepackage{xcolor}
\usepackage{nicefrac}
\usepackage{orcidlink}

\usepackage[]{graphicx}\usepackage[]{xcolor}
\makeatletter
\def\maxwidth{ %
  \ifdim\Gin@nat@width>\linewidth
    \linewidth
  \else
    \Gin@nat@width
  \fi
}
\makeatother

\definecolor{fgcolor}{rgb}{0.345, 0.345, 0.345}
\newcommand{\hlsng}[1]{\textcolor[rgb]{0.192,0.494,0.8}{#1}}%
\newcommand{\hldef}[1]{\textcolor[rgb]{0.345,0.345,0.345}{#1}}%
\newcommand{\hlkwc}[1]{\textcolor[rgb]{0.333,0.667,0.333}{#1}}%
\newcommand{\hlkwd}[1]{\textcolor[rgb]{0.737,0.353,0.396}{\textbf{#1}}}%

\usepackage{framed}
\makeatletter
\newenvironment{kframe}{%
 \def\at@end@of@kframe{}%
 \ifinner\ifhmode%
  \def\at@end@of@kframe{\end{minipage}}%
  \begin{minipage}{\columnwidth}%
 \fi\fi%
 \def\FrameCommand##1{\hskip\@totalleftmargin \hskip-\fboxsep
 \colorbox{shadecolor}{##1}\hskip-\fboxsep
     \hskip-\linewidth \hskip-\@totalleftmargin \hskip\columnwidth}%
 \MakeFramed {\advance\hsize-\width
   \@totalleftmargin\z@ \linewidth\hsize
   \@setminipage}}%
 {\par\unskip\endMakeFramed%
 \at@end@of@kframe}
\makeatother

\definecolor{shadecolor}{rgb}{.97, .97, .97}
\definecolor{messagecolor}{rgb}{0, 0, 0}
\definecolor{warningcolor}{rgb}{1, 0, 1}
\definecolor{errorcolor}{rgb}{1, 0, 0}
\newenvironment{knitrout}{}{} 

\usepackage{alltt}

\newcommand{\AUC}{\text{AUC}}
\newcommand{\ROC}{\text{ROC}}

\newcommand{\Sec}{\text{Se}^\mathbb{C}}
\newcommand{\Spd}{\text{Sp}^\mathbb{D}}
\newcommand{\Sps}{\mathrm{Sp}^{\overline{\mathbb{D}}}}
\newcommand{\Sei}{\text{Se}^\mathbb{I}}
\newcommand{\ROCcd}{\ROC^{\mathbb{C}/\mathbb{D}}}

\newcommand{\rY}{Y}
\newcommand{\rT}{T}

\newcommand{\rX}{\mX}

\newcommand{\rx}{\xvec}

\newcommand{\FY}{F_Y}
\newcommand{\FT}{F_T}
\newcommand{\lpY}{\rx^\top \betavec_Y}
\newcommand{\lpT}{\rx^\top \betavec_T}

\newcommand{\hY}{h_\rY}
\newcommand{\hT}{h_\rT}

\newcommand{\basisy}{\bvec}

\newcommand{\parm}{\varthetavec}

\newcommand{\tlw}{\underline{t}}
\newcommand{\tup}{\overline{t}}
\newcommand{\ylw}{\underline{y}}
\newcommand{\yup}{\overline{y}}

\newcommand{\RR}{\mathbb{R}}

\usepackage{dsfont}

 \DeclareMathOperator{\expit}{expit}

\def \bvec {\text{\boldmath$b$}}

    \def \mO {\text{\boldmath$O$}}

\def \xvec {\text{\boldmath$x$}}    \def \mX {\text{\boldmath$X$}}
    
    \def \mZ {\text{\boldmath$Z$}}

\def \betavec         {\text{\boldmath$\beta$}}
\def \gammavec        {\text{\boldmath$\gamma$}}

\def \thetavec        {\text{\boldmath$\theta$}}
\def \varthetavec     {\text{\boldmath$\vartheta$}}

\def \mSigma   {\mathbf{\Sigma}}

\def \nullvec {\mathbf{0}}

\newcommand{\indep}{\perp \!\!\! \perp}

\newcommand\sbullet[1][.5]{\mathbin{\vcenter{\hbox{\scalebox{#1}{$\bullet$}}}}}

\newcommand{\AYcite}[2]{\cite{#2}}

\author{Ainesh Sewak\orcidlink{0000-0003-1858-9987} \\ Universit\"at Bern
	\And Vanda In{\'a}cio\orcidlink{0000-0001-8084-1616} \\ University of Edinburgh
	\AND Joanne Wuu\orcidlink{0009-0005-9643-9855} \\ University of Miami
	\And Michael Benatar\orcidlink{0000-0003-4241-5135} \\ University of Miami
	\And Torsten Hothorn\orcidlink{0000-0001-8301-0471} \\ Universit\"at Z\"urich}
\Plainauthor{Ainesh Sewak, Vanda In{\'a}cio, Joanne Wuu, Michael Benatar, Torsten Hothorn}

\title{Likelihood-based Modeling of Covariate-Specific Time-Dependent ROC Curves}
\Shorttitle{Covariate-Specific Time-Dependent ROC Modeling}
\Plaintitle{}

\Abstract{
	Identifying reliable biomarkers for predicting clinical events in longitudinal studies is important for accurate disease prognosis and for guiding development of new treatments. However, prognostic studies are often observational, making it difficult to account for patient heterogeneity. In amyotrophic lateral sclerosis (ALS), factors such as age, site of onset and genetic status influence both survival and biomarker levels, yet their impact on the prognostic accuracy of biomarkers over time remains unclear. While time-dependent receiver operating characteristic methods have been developed to handle censored time-to-event outcomes, most do not adjust for covariates. 
To address this, we propose the nonparanormal prognostic biomarker framework, which models the joint distribution of the biomarker and event time while accounting for covariates. This allows estimation of covariate-specific time-dependent ROC curves and related summary measures.
We apply the NPB framework to evaluate serum neurofilament light as a prognostic biomarker in ALS, showing that its accuracy varies over time and with patient characteristics. By capturing these covariate-specific effects, the NPB framework supports more targeted risk stratification and  can potentially improve the design of clinical trials for new ALS treatments.
 }

\Plainkeywords{time-dependent ROC analysis, covariates, prognostic biomarkers, amyotrophic lateral sclerosis, censoring}

\Address{
	Ainesh Sewak \\
	Department of Clinical Research \\
	Universit\"at Bern \\
	Mittelstrasse 43, CH-3010 Bern, Switzerland\\
	Email: \texttt{Ainesh.Sewak@unibe.ch} \\
	
	Vanda In{\'a}cio \\
	School of Mathematics \\
	University of Edinburgh \\
	Peter Guthrie Tait Road, EH9 3FD Edinburgh, United Kingdom \\
	Email: \texttt{Vanda.Inacio@ed.ac.uk} \\
	
	Joanne Wuu, Michael Benatar \\
	Department of Neurology \\
	University of Miami Miller School of Medicine\\
	1120 NW 14th Street, Suite 1300, Miami, FL, 33136, United States \\
	Email: \texttt{MBenatar@med.miami.edu, JWuu@med.miami.edu} \\

	Torsten Hothorn\\
	Institut f\"ur Epidemiologie, Biostatistik und Pr\"avention \\
	Universit\"at Z\"urich \\
	Hirschengraben 84, CH-8001 Z\"urich, Switzerland \\
	Email: \texttt{Torsten.Hothorn@R-project.org} \\

}

\begin{document}

\section{Introduction}

Prognostic biomarkers are essential tools for predicting clinical events. They can enable more effective patient stratification and improved design of clinical trials. As the discovery of new biomarkers accelerates, so too does the need for methods to evaluate their predictive accuracy, particularly in the presence of patient heterogeneity. 

Two key considerations in prognostic biomarker evaluation is that a marker’s accuracy may vary over time and across subgroups defined by patient-level covariates. Understanding these is important for assessing the biomarker’s clinical utility. Unlike diagnostic studies, where disease status and biomarkers are measured concurrently, prognostic studies evaluate a baseline biomarker’s ability to predict future outcomes, often subject to right- or interval-censoring.

Unlike randomized trials, prognostic biomarker studies are typically observational, meaning treatment or exposure groups are not assigned at random~\citep{freidlin2010randomized}. As a result, covariate imbalance and confounding are common, making it difficult to disentangle the prognostic value of the biomarker from that of patient characteristics. This challenge is particularly relevant in diseases like amyotrophic lateral sclerosis (ALS), where age, disease onset site and genetic status can influence both survival and biomarker levels.

The most common approach to handle this in medical research is to use the hazard ratio from a proportional hazards model as a summary of prognostic performance. However, association measures such as hazard ratios do not directly quantify the accuracy of a biomarker to discriminate patients at higher risk of experiencing the event~\citep{pepe2004limitations, bansal2019comparison}. Moreover, the proportional hazards assumption is frequently violated in practice, potentially leading to misleading inference~\citep{stensrud2025invited}.

To address these limitations, time-dependent extensions of classical accuracy metrics such as sensitivity, specificity, the receiver operating characteristic (ROC) curve and the area under the curve (AUC) have been developed~\citep{slate2000statistical, heagerty2005survival, cai2006sensitivity}. In this work, we focus on the cumulative sensitivity and dynamic specificity framework, which is widely adopted in clinical research~\citep{lambert2016summary}. A large body of work exists on time-dependent ROC analysis, including nonparametric and semiparametric estimators. For a review, see \AYcite{Blanche et al.; Kamarudin et al.}{blanche2013review, kamarudin2017time}.

However, comparatively fewer methods account for covariates when assessing prognostic accuracy, despite growing evidence that ignoring patient heterogeneity can bias ROC-based estimates~\citep{janes2008adjusting}. Three main classes of methods have been proposed for covariate adjustment in time-dependent ROC analysis:
(1) model-based approaches that rely on proportional hazards models for event times given the biomarker and covariates~\citep{song2008semiparametric, li2015estimation}, which are sensitive to the proportional hazards model assumption;
(2) fully nonparametric methods using kernel smoothing~\citep{rodriguez2016nonparametric}, which are limited to a single continuous covariate and exhibit high variance;
and (3) more recent strategies based on covariate standardization or marginalization~\citep{le2018standardized, dey2023inference, jiang2024addressing}, which provide population-averaged rather than covariate-specific accuracy estimates.

In this paper, we propose a modeling framework for covariate-specific, time-dependent ROC analysis that addresses several key limitations of existing methods. Our approach combines semiparametric marginal transformation models for the biomarker and event times, which incorporate covariate effects and are linked via a copula to capture their joint dependence. The model is efficiently parameterized, making it well-suited for small to moderate sample sizes that are common in prognostic biomarker studies. Unlike fully nonparametric approaches, our framework is more efficient, accommodates multiple covariates and allows for both covariate-specific and population-averaged accuracy estimates. A major advantage is its ability to handle both right- and interval-censored outcomes within a unified likelihood framework, avoiding the need for separate estimators~\citep{beyene2020smoothed, beyene2022time}. While the Gaussian copula imposes a structured form of dependence, it enables tractable estimation and facilitates meaningful extensions, such as modeling detection limits for biomarkers and covariate-dependent censoring, that are difficult to implement using kernel-based or fully nonparametric methods.

The layout of the paper is as follows. We begin with our motivation in Section~\ref{sec:data}, using biomarker and event time data from a longitudinal study on amyotrophic lateral sclerosis (ALS).
Section~\ref{sec:eval} introduces key concepts, including sensitivity, specificity and the covariate-specific cumulative-dynamic ROC curve. 
In Section~\ref{sec:npbm}, we present the technical details of our model, including parameterization and inference.
In Section~\ref{sec:extension}, we explore potential extensions of our approach and detail a technique for model assessment.
Section~\ref{sec:simulation} provides results from a simulation study with a detailed comparison to existing methods. 
In Section~\ref{sec:application}, we apply our method to the ALS study data.
We offer concluding remarks in Section~\ref{sec:conclusion}. 
\section{Prognostic biomarkers in ALS}
\label{sec:data}

ALS is a progressive neurodegenerative disease that causes degeneration of nerve cells, leading to muscle weakness, paralysis and respiratory failure. With a median survival duration of 20 to 48 months after symptom onset, ALS currently has no cure and available treatments primarily aim to manage the symptoms and slow disease progression \citep{feldman2022amyotrophic}. Identifying reliable prognostic biomarkers for ALS are essential for improving clinical trial design~\citep{kiernan2021improving, benatar2024prognostic}.
Biomarkers may be utilized in trial designs to stratify patients into homogeneous subgroups, serve as eligibility criteria to define appropriate study populations and act as baseline covariates to adjust for patient heterogeneity. Together, these applications reduce sample size requirements, enable more cost-effective studies, and accelerate the development of effective treatments~\citep{taga2018current}.

\subsection{Longitudinal ALS cohort}

The methods developed in this article are motivated by a prospective
longitudinal study on the prognostic utility of neurofilament biomarkers for
amyotrophic lateral sclerosis (ALS), as described
by~\AYcite{Benatar et al.}{benatar2020validation}.  One of the outcomes of the study was
survival duration, defined as the time from the baseline visit to either
permanent assisted ventilation, tracheostomy, or death.  Patients who did
not experience any of these events during the follow-up period were
right-censored.  Our study analyzed a subset of $260$ patients enrolled across
multiple clinical sites through the Clinical Research in ALS and Related
Disorders for Therapeutic Development (CReATe) Consortium’s
Phenotype-Genotype-Biomarker study (registered at clinicaltrials.gov: NCT02327845), representing the data available for
biomarker analysis at the time of a prior publication~\citep{benatar2020validation}.  
Patients included in the present study comprised $229$ patients with ALS, $11$ with progressive muscle
atrophy (PMA), and $20$ with primary lateral sclerosis.  To discover and
validate biomarkers, all patients underwent systematic follow-ups with
standardized clinical evaluations and provided biological samples every $3$ to
$6$ months.  For our analysis, we included patients with ALS or PMA.  After
excluding $12$ patients with missing values in one or more covariates, the
final sample comprised $N = 218$ patients.

\begin{figure}[t!]
	\centering
	\includegraphics[width=0.85\linewidth]{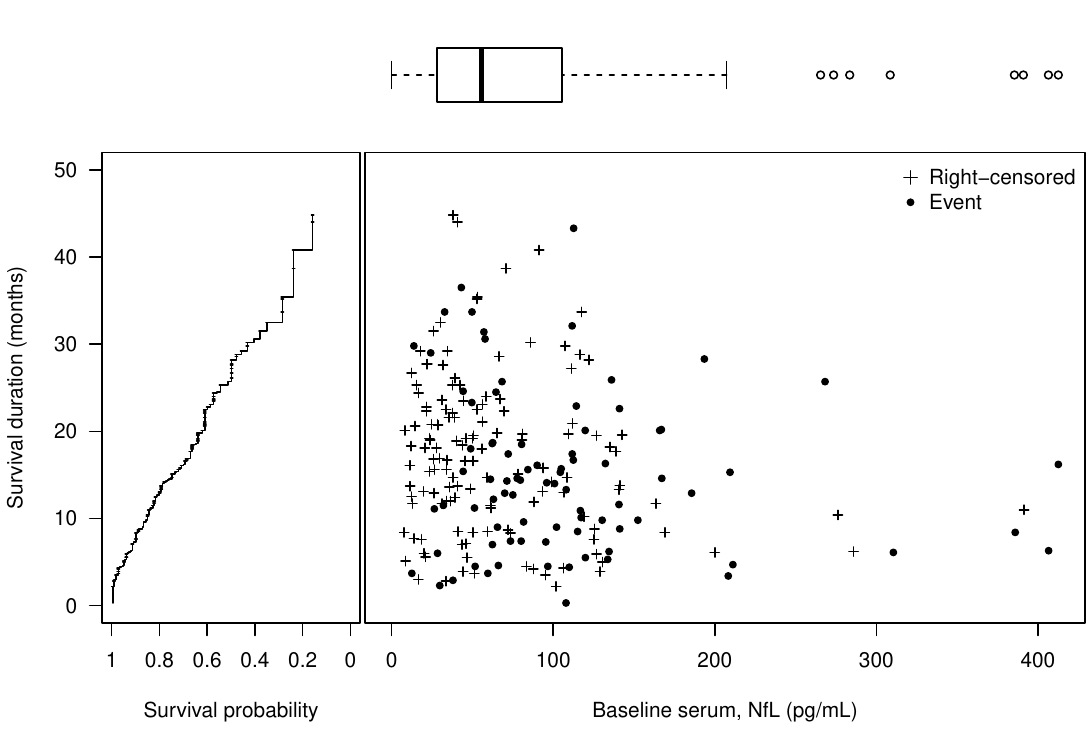} 
	\caption{Scatter plot of the observed baseline serum neurofilament light~(NfL) concentration (pg/mL) and survival duration (in months since baseline). Right censored subjects are indicated with~``+'' and subjects who reached the endpoint are marked by~``$\sbullet[0.5]$''.}
	\label{fig:scatter}
\end{figure}

The biomarker of interest in our study is serum neurofilament light (NfL) concentration. 
Baseline serum NfL concentration has been shown to differentiate ALS patients from healthy controls and is associated with both survival outcomes and disease progression in ALS \citep{lu2015neurofilament, benatar2024prognostic}.
Figure~\ref{fig:scatter} illustrates the observed relationship between baseline NfL and survival duration in the present study, showing that higher baseline NfL levels are generally associated with shorter survival. However, this association is not strictly linear and any evaluation of prognostic accuracy must account for the right-censored data in the study. Furthermore, the prognostic accuracy of NfL may depend on the time horizon of prediction and could be influenced by patient characteristics.

\subsection{Challenges in assessing prognostic accuracy} 
Evaluating the prognostic accuracy of ALS biomarkers in observational studies is challenging because of the risk of confounding. Confounding arises when covariates such as age are associated with both biomarker levels and disease prognosis. For example, older patients tend to have elevated NfL levels and face an increased risk of ALS-related mortality, potentially distorting the accuracy of NfL as a prognostic marker. While factors such as aging and site of onset are known to be associated with NfL levels, their impact on NfL’s prognostic accuracy across different time horizons remains unclear~\citep{koini2021factors, benatar2023neurofilament}.

Stratifying by covariates can help address confounding, but this approach is limited to a small number of categorical covariates and often results in small sample sizes in each group, limiting the ability to draw statistically efficient conclusions. In ALS, where large datasets are scarce, proportional hazards models are commonly used to estimate an overall hazard ratio. Such regression models account for confounding and can give an indication of the prognostic utility of NfL by quantifying its association with survival duration. However, they do not fully capture how NfL’s prognostic accuracy changes across different time horizons. Additionally, these models assume that the relationship between NfL, covariates, and survival duration remains proportional over time. This is a condition that may not hold in a progressive disease like ALS, where risk dynamics evolve \citep{stensrud2025invited}.

To address these limitations, covariate-specific sensitivity and specificity can be used to assess a biomarker’s ability to predict survival. These measures form the foundation of ROC analysis and provide a framework for evaluating prognostic accuracy across different time horizons.


\section{Evaluation of prognostic biomarkers}
\label{sec:eval}
We first provide a brief overview of evaluating the accuracy of a continuous biomarker $Y$ in predicting a time-constant binary outcome $D$, where $D = 0$ represents ``controls'' and $D = 1$ represents ``cases''. Assuming higher values of $Y$ are more indicative of disease, a threshold $c \in \RR$ can be used to classify subjects as cases if their biomarker value exceeds $c$, and as controls otherwise. Sensitivity is the correct classification probability for cases, defined as $\text{Se}(c) = \Prob(Y > c \mid D = 1)$, while specificity represents the correct classification probability for controls, $\text{Sp}(c) = \Prob(Y < c \mid D = 0)$. For any given biomarker, there is a trade-off between sensitivity and specificity, as increasing one decreases the other. This trade-off is determined by the choice of $c$. The ROC curve is a graphical tool that plots sensitivity against $1-\text{specificity}$ for all possible threshold values of~$c$ and provides a measure of the biomarker’s diagnostic accuracy.

Most diseases do not present binary outcomes. Instead, they often involve longitudinal outcomes, such as survival duration
for ALS patients. To accommodate this, various extensions have been developed to measure time-dependent sensitivity and specificity \citep{heagerty2005survival}. We focus on the cumulative sensitivity and dynamic specificity because our interest lies in evaluating the accuracy of prognostic ALS biomarkers at specific time points. Additionally, both the biomarker and survival duration might depend on a set of covariates $\rX = (X_1,\dots,X_p)$ such as age, sex, genetic factors and disease progression which need to be taken into account.

For the event time $T$, we define the covariate-specific cumulative sensitivity and dynamic specificity as
\begin{align*}	
	\Sec_t(c \mid \rx) &= \Prob(Y > c \mid T \leq t, \rX = \rx) \\
	\Spd_t(c \mid \rx) &= \Prob(Y \leq c \mid T > t, \rX = \rx).
\end{align*}
In this context, \emph{cumulative} sensitivity considers any subject who experiences the event before a time point of interest $t$ as a case (i.e., $T \leq t$, or conceptually, $D(t) = I(T \leq t) = 1$), while \emph{dynamic} specificity defines a control as any subject who remains event-free at time $t$ (i.e., $T > t$, or $D(t) = 0$). For a fixed time $t$, the entire population is thus classified into cases (those who have experienced the event) and controls (those still at risk).

These definitions lead to the \emph{cumulative-dynamic} ROC curve, which assesses the biomarker’s prognostic accuracy for a given time horizon. This enables evaluation of how effectively a baseline marker can differentiate between subjects who will experience the event of interest and those who will not within a specified follow-up interval. Each time horizon of interest may yield a different accuracy, reflecting the dynamic nature of time-to-event data. 
The covariate-specific cumulative-dynamic ROC curve is given by
\begin{align*}
	\ROCcd_t(p \mid \rx) = \Sec_t \left( [1 - \Spd_t]^{-1}(p \mid \rx) \mid \rx \right), \; \text{for} \; p \in [0, 1],
\end{align*}
where the function $[1 - \Spd_t]^{-1}( p ) = \inf \left\{c \in \RR : 1- \Spd_t(c) \leq p  \right\}$ exists. These definitions allow for assessing a biomarker's prognostic accuracy for subjects with shared characteristics $\rX = \rx$. 

The area under the cumulative-dynamic ROC curve (AUC) is the most widely used summary measure of the ROC curve and represents the probability that a subject with a higher biomarker value experienced the event earlier. The corresponding covariate-specific time-dependent AUC is defined as
\begin{align*}
	\AUC_t(\rx) = \int_0^1 \ROCcd_t(p \mid \rx) \, d p
	= \Prob(Y_i > Y_j \mid T_i \leq t, T_j > t, \rX_i = \rX_j = \rx)
\end{align*}
for two distinct subjects $i$ and $j$. Other measures, such as the Youden Index and optimal cut-off values, can also be derived once the ROC curve is obtained. 
Although this paper focuses on the cumulative-dynamic ROC curve, the methods discussed can estimate other forms of time-dependent ROC curves in the same statistical framework as well. Additional details for these approaches are provided in 
Supplementary Material~\ref{app:alt_roc}. 

\section{The nonparanormal prognostic biomarker model}
\label{sec:npbm}

We propose the nonparanormal prognostic biomarker (NPB) model as a flexible framework for covariate-specific, time-dependent ROC analysis. The NPB model comprises two main components: (1) a joint model that captures the dependence between the biomarker and event time, and (2) marginal transformation models for each variable that incorporate covariate effects. The name ``nonparanormal'' reflects the structure of the likelihood, which combines these two components into a unified framework.

Using Bayes' theorem, for a fixed time $t$, the cumulative sensitivity can be expressed as
\begin{align*}
	\Sec_t(c \mid \rx) = \frac{\Prob(Y > c, T \leq t \mid \rX = \rx)}{\Prob(T \leq t \mid \rX = \rx)}
	= \frac{\FT(t \mid \rx) - F_{Y, T}(c, t \mid \rx)}{\FT(t \mid \rx)},
\end{align*}
where $\FT(t \mid \rx)$ is the marginal cumulative distribution function (CDF) of the event time conditioned on the covariates $\rx$ and $F_{Y, T}(y, t \mid \rx)$ is the joint conditional CDF of the biomarker and event time. Similarly, the dynamic specificity at time $t$ is given by
\begin{align*}
	\Spd_t(c \mid \rx) = \frac{\Prob(Y \leq c, T > t \mid \rX = \rx)}{\Prob(T > t \mid \rX = \rx)}
	= \frac{\FY(c \mid \rx) - F_{Y, T}(c, t \mid \rx)}{1 - \FT(t \mid \rx)},
\end{align*}
where  $\FY(c \mid \rx)$ is the marginal CDF of the biomarker. Notice that if the joint conditional distribution of $(Y, T)$ given $\rX = \rx$ is known, all terms in these expressions can be derived. Further, we can calculate cumulative sensitivity and dynamic specificity for a range of cut-off points $c \in \RR$ and plot the corresponding ROC curve for a specific time $t$ conditional on covariate values $\rx$. The time-dependent covariate-specific AUC can then be obtained by numerically integrating the corresponding ROC curve.

In this paper, we focus on flexibly modeling the joint conditional distribution $F_{Y,T}(y, t \mid \rx)$ of the biomarker and event time given $\rX = \rx$. This approach allows us to calculate model-based covariate-specific cumulative sensitivity, dynamic specificity, and construct time-dependent covariate-specific ROC curves and corresponding summary metrics such as $\AUC_t(\rx)$. Next, we detail the components of the model followed by a likelihood-based estimation approach that can accommodate both right and interval censoring schemes.

%

\subsection{Joint model specification}

Semiparametric regression models that incorporate covariate effects are well-established for modeling either the distribution of a biomarker~\citep{pepe1997regression} or the distribution of event times~\citep{cox1972regression}, both of which may be subject to censoring. However, jointly modeling the conditional distribution of the biomarker and the event time poses additional challenges, particularly in capturing their dependence structure in a way that accommodates covariates and censoring.

To address this, we adopt a nonparanormal copula model~\citep{liu2009nonparanormal}, which links semiparametric marginal models for the biomarker and event time via a Gaussian copula. Specifically, we model the conditional joint distribution of $(Y, T) \mid \rX = \rx$ as
\begin{align*}
	F_{Y, T}(y, t \mid \rx) &= \Prob(Y \leq y, T \leq t \mid \rX = \rx) \\  
	&= C_{\rho}(F_Y(y \mid \rx), F_T(t \mid \rx)) \\
	&=\Phi_{\rho}(\Phi^{-1}(F_Y(y \mid \rx)), \Phi^{-1}(F_T(t \mid \rx)))
\end{align*}
where $\Phi$ denotes the CDF of a standard normal distribution and
$\Phi_{\rho}(z_1, z_2)$ is the CDF of the bivariate standard normal distribution
with correlation parameter $\rho \in [-1, 1]$.
The copula $C_{\rho}$ captures the dependence structure between the biomarker and event time, leaving their marginal distributions unspecified. We define monotonically nondecreasing marginal transformation functions as
\begin{align*}
	\hY(y \mid \rx) = \Phi^{-1}(\FY(y \mid \rx)) \; \text{and} \; \hT(t \mid \rx) = \Phi^{-1}(\FT(t \mid \rx)),
\end{align*}
which map the conditional marginals to the standard normal scale. Under this model, the latent variables follow a bivariate normal distribution
\begin{align*}
	\mZ = \left(\hY(Y \mid \rx), \hT(T \mid \rx)  \right)^\top \sim N_2\left(\nullvec, \begin{pmatrix} 1 & \rho \\ \rho & 1 \end{pmatrix}\right).
\end{align*}
This transformation simplifies inference while allowing flexible specification of each marginal via semiparametric transformation models. In practice, the marginal models for $Y$ and $T$ may depend on different covariates. We assume the same covariate set for notational simplicity.

While we focus on the Gaussian copula due to its computational tractability, particularly the availability of closed-form likelihood and score functions, our framework can accommodate other parametric copulas if needed. In Section~\ref{sec:extension}, we also describe how to extend the model to allow for covariate-specific correlation, which relaxes the assumption of constant dependence across the population.

\subsection{Marginal model parameterization}
We model the marginal distributions semiparametrically using probit linear transformation models
\begin{align*}
	\FY(y \mid \rx) &= \Phi\left(\hY(y) - \lpY \right) \; \text{and} \; \FT(t \mid \rx) = \Phi\left(\hT(t) - \lpT \right),
\end{align*}
where $\hY$ and $\hT$ are baseline transformation functions that map the biomarker $Y$ and event time $T$ to the latent normal scale, and $\betavec_Y$ and $\betavec_T$ are the respective linear covariate effects. Under this specification, the joint conditional CDF of $(Y,T) \mid \rX=\rx$ is given by
\begin{align}
	F_{Y, T}(y, t \mid \rx) &= \Phi_{\rho}\left(\hY(y) - \lpY, \hT(t) - \lpT \right),
	\label{eq:cdf}
\end{align}
and the corresponding joint conditional density function is
\begin{align}
	f_{Y, T}(y, t \mid \rx) = \phi_{\rho} \left(\hY(y) - \lpY, \hT(t) -\lpT \right) \hY'(y) \hT'(t).
	\label{eq:dens}
\end{align}
We adopt the probit link here for notational clarity, but other choices are compatible within this framework. For instance, the transformation functions can be specified to correspond to proportional hazards or proportional odds models. These alternatives are discussed in detail in Section~\ref{sec:extension}.

We model the baseline transformation function for the biomarker as a smooth, monotonically nondecreasing function of $y$, defined as
\begin{align*}
	\hY(y \mid \parm_Y) = \basisy(y)^\top \parm_Y = \sum_{m=0}^M \vartheta_{Y,m} b_m(y),
\end{align*}
where $\basisy(y) = (b_0(y),\dots, b_M(y))^\top$ is a vector of $M + 1$ basis functions with coefficients $\parm_Y \in \RR^{M+1}$~\citep{hothorn2018most}. We use Bernstein polynomials as the basis functions, as they provide a flexible yet structured way to approximate any continuous function on a bounded interval~\citep{farouki2012bernstein}. The degree $M$ controls the smoothness of the transformation; for example, $M = 1$ corresponds to a linear transformation, while in practice we find that $M=6$ is sufficient for capturing most empirical biomarker distributions~\citep{sewak2023estimating}.

To ensure that $\hY$ is nondecreasing, we impose monotonicity constraints on the coefficients $\vartheta_{Y,m} \leq \vartheta_{Y, m+1}$ for $m = 0, \dots, M-1$. The Bernstein basis polynomial of order $M$ on the interval $[l, u]$ is given by
\begin{align*}
	b_m(y) = \binom{M}{m} \tilde{y}^m (1 - \tilde{y})^{M - m}, \; m = 0,\dots,M,
\end{align*}
where the observations are rescaled to the unit interval via $\tilde{y} = \frac{y - l}{u - l}$. The baseline transformation function for survival time is parameterized in the same manner, with its own coefficient vector denoted by~$\parm_T$.

\subsection{Likelihood inference}
The complete set of parameters for our model is denoted as $\thetavec = \left(\parm_Y^\top, \parm_T^\top, \betavec_Y^\top, \betavec_T^\top, \rho \right)^\top$. The likelihood in our model corresponds to the
``mixed nonparanormal likelihood'' with smooth parameterization of the marginal transformation functions \citep{hothorn2024on}. 
This mixed likelihood structure accommodates both censored and continuous
responses. It results in a ``flow nonparanormal likelihood'' for continuous variables and
an interval-censored likelihood defined by the conditional distribution of 
censored event time observations.
In the bivariate case considered here, our model accounts for two main types of observations: exact and interval-censored. Let $N_e$ denote the number of observations with exact event times and $N_c$ those with interval-censored events, for a total of $N = N_e + N_c$ observations.

For exact (uncensored) observations, we observe $O_i = (y_i, t_i, \rx_i)$ for $i = 1,\dots, N_e$.  The likelihood contribution is given by the joint conditional density
\begin{align*}
	\ell_e(\thetavec \mid O_i) = f_{Y,T}(y_i, t_i \mid \rx_i, \thetavec)
\end{align*}
as defined in Equation~\ref{eq:dens}. 

For interval-censored event times,
where biomarker values are observed exactly and the event time falls in a known interval $ (\tlw_i, \tup_i)$, the observation is $O_i = (y_i, \tlw_i, \tup_i, \rx_i)$ for $i = N_e + 1,\dots, N$. Assuming conditional independence, the likelihood contribution is
\begin{align*}
	\ell_c(\thetavec \mid O_i)
	&= \int_{\hT(\tlw_i \mid \vartheta_T)}^{\hT(\tup_i \mid \vartheta_T)} \phi_{\rho} \left(\hY(y_i \mid \vartheta_Y) - \rx_i^\top \betavec_Y, z_T - \rx_i^\top \betavec_T \right) \hY'(y_i) dz_T,
\end{align*}
where $\phi_{\rho}$ is the bivariate normal density with correlation $\rho$ and the integral represents the probability mass over the censored time interval. 
Right-censored observations are handled as a special case  of this likelihood by setting $\tup_i = \infty$. The full log-likelihood is then constructed as
\begin{align*}
	\ell(\thetavec \mid \mO) = \sum_{i = 1}^{N_e} \log(\ell_e(\thetavec \mid O_i)) + \sum_{i = N_e + 1}^{N} \log(\ell_c(\thetavec \mid O_i)),
\end{align*}
where $\mO = \{O_1,\dots, O_{N}\}$ denotes the complete dataset.

Maximum likelihood estimation of $\thetavec$ is carried out via constrained optimization, requiring repeated evaluation of multivariate normal probabilities. These are efficiently computed using randomized quasi–Monte Carlo integration methods~\citep{genz1992numerical} and implemented in the \pkg{mvtnorm} package~\citep{multivariate2024hothorn}. The corresponding score functions are available in closed form and described in detail elsewhere~\citep{hothorn2024on}. A practical advantage of the proposed framework is that the estimation of the marginal distributions can be performed separately from the dependence structure. This allows for efficient and numerically stable initialization of parameters through low-dimensional, convex optimization problems, which facilitates reliable convergence when fitting the full joint model.

\section{Extensions and model assessment}
\label{sec:extension}

In this section, we outline several extensions of the NPB framework that increase its applicability in more complex settings. These include handling limits of detection in biomarker measurements, accommodating informative or non-independent censoring, modeling covariate-specific correlation structures and incorporating nonlinear covariate effects. While these features were not applied in the ALS case study, the NPB framework can support them. Importantly, all of these extensions are already implemented in the \pkg{tram} package in \proglang{R}. We also present a graphical model assessment strategy based on PIT transformations to evaluate model calibration.

\subsection{Limit of detection in biomarkers}
Biomarker measurements are often affected by random errors and limits of detection issues. Ignoring of such a data characteristic can reduce the biomarker's prognostic accuracy. Statistically, this is a form of interval censoring where measurements below the detection limit are left-censored and those above the detection limit are right-censored. In the NPB framework, the likelihood can be adjusted to account for this censoring in the biomarker data. For independent interval-censored biomarker measurements and interval-censored event times, we observe $O_i = (\ylw_i, \yup_i, \tlw_i, \tup_i, \rx_i)$ and the likelihood contribution is expressed as
\begin{align*}
	\ell_c(\thetavec \mid O_i) 
	&= \int_{\hY(\ylw_i \mid \vartheta_Y)}^{\hY(\yup_i \mid \vartheta_Y)} \int_{\hT(\tlw_i \mid \vartheta_T)}^{\hT(\tup_i \mid \vartheta_T)} \phi_{\rho} \left(z_Y - \rx_i^\top \betavec_Y, z_T - \rx_i^\top \betavec_T \right) \, d z_Y \, d z_T.
\end{align*}
In this formulation, lower detection limits are treated as a special case where $\ylw_i = -\infty$ and upper detection limits are treated as $\yup_i = \infty$.
Inference can subsequently proceed as detailed previously and the resulting estimates would not suffer from the negative bias of ignoring the true nature of the biomarker measurement.

\subsection{Dependent censoring}

In the methods presented above, we assumed conditionally independent censoring. That is, once the covariates $\rX$ are known, the event time
$T$ and censoring time $C$ are independent, i.e., $T \indep C \mid \rX = \rx$. However, this assumption may not hold in practice. Event time can influence censoring time and censoring may depend on biomarker levels. For example, in ALS, elevated levels of NfL might be associated with a higher risk of death, leading to study dropout as patients become too sick for follow-up.

To address this, the NPB framework can be extended to model the joint relationship between $Y$, $T$, and $C$ given $\rX = \rx$, allowing for non-independent censoring. Under certain conditions, such a model is identifiable when the marginal distributions of $T$ and $C$ are parametrically specified, for example, as log-normal or Weibull distributions~\citep{czado2023dependent}. Alternatively, the event time can be modeled semiparametrically with a Cox proportional hazards model while censoring times follow a Weibull model \citep{deresa2024copula}.
From any such trivariate model, we can derive the bivariate distribution of $(Y, T) \mid \rX = \rx$ using the Gaussian copula, now parameterized by a correlation matrix rather than a single parameter. The multivariate normal distribution's properties allow us to subset the rows and columns of the correlation matrix related to $Y$ and $T$, and proceed with calculating cumulative sensitivity and dynamic specificity as in the original model.

\subsection{Covariate-specific correlation}
In some medical conditions, the correlation between a biomarker and event time can vary based on covariates. For example, in ALS, the correlation between NfL levels and survival duration may depend on covariates such as age or disease onset site. Younger patients with limb-onset ALS might exhibit a weaker correlation between NfL and survival, whereas older patients or those with bulbar-onset ALS may show a stronger correlation due to faster disease progression.

In the NPB framework, the correlation matrix $\mSigma$ can vary with the covariates $\rx$, allowing the dependence structure between $Y$ and $T$ to change as a function of $\rx$. For the bivariate case, the correlation between $Y$ and $T$ for a given set of covariates $\rx$ is given by
\begin{align*}
	\rho(\rx) = \dfrac{-\lambda(\rx)}{\sqrt{\lambda(\rx)^2 + 1}},
\end{align*}
where $\lambda(\rx) = \alpha + \rx^\top \gammavec$ is one possible parameterization of the coefficient of the inverse Cholesky factor~\citep{klein2022multivariate, barratt2023covariance}. This extension allows the model to capture subgroup-specific relationships between the biomarker and event time.

\subsection{Alternative marginal models}
The assumption of probit semiparametric transformation models can also be relaxed. An alternative approach is to use a continuous odds logistic regression model for the biomarker, known for its robustness properties \citep{harrell2015ordinal}, alongside a proportional hazards model for the event time
\begin{align*}
	\FY(y \mid \rx) &= \expit\left(\hY(y) - \lpY \right) \; \text{and} \; \FT(t \mid \rx) = 1 - \exp\left( - \exp \left( \hT(t) - \lpT \right) \right).
\end{align*}
In general, any absolutely continuous cumulative distribution function with a log-concave density can be substituted for the probit link function, providing flexibility in marginal model choice.

\subsection{Nonlinear covariate modeling}

The linear predictor in our marginal models may be insufficient to fully capture the effects of covariates in some cases. To address this, we can incorporate covariates directly into the NPB framework by estimating the joint distribution of $(Y, T, \rX)$
using the parameterizations and estimation procedures described earlier~\citep{hothorn2024on}. This aligns to the approach of \AYcite{Rodr\'{\i}guez-\'Alvarez et al.}{rodriguez2016nonparametric} but can be extended to incorporate multiple covariates. Continuous covariates can be parameterized using any polynomial or spline basis. To generate covariate-specific time-dependent ROC curves, we can determine the
conditional distribution of $(Y, T) \mid \rX = \rx$ from the Gaussian copula.

\subsection{Model assessment}

Model assessment is important for evaluating the adequacy of both the marginal distributions and the joint dependence structure. A key tool for assessing calibration is the probability integral transform (PIT)~\citep{dawid1984present}.
For a continuous random variable $Z$ with true cumulative distribution function $F_Z$, the transformed value $F_Z(Z)$ follows a standard uniform distribution $U(0, 1)$. We can substitute the estimated CDF $\hat{F}_Z$ and compute $\hat{F}_Z(Z_i)$ on the observed data. If the model is correctly specified, the PIT values should approximately follow an uniform distribution. This can be assessed visually using a histogram or a quantile-quantile (QQ) plot~\citep{gneiting2007probabilistic}. Systematic deviations from uniformity suggest miscalibration or misspecification in the modeled distribution.
	
The PIT can be extended to the multivariate setting via the Rosenblatt transformation~\citep{rosenblatt1952remarks}. For the bivariate vector of biomarker and event time $(Y, T)$, we define the transformed variables as
\begin{align*}
	U_1 = F_T(T), \quad U_2 = F_{Y \mid T}(Y \mid T)
\end{align*}
where $U_1$ and $U_2$ are independent and uniformly distributed on $[0, 1]$ if the joint model is correctly specified. In our framework, these conditional distributions are univariate Gaussian and the PIT quantities can be computed directly from the fitted model parameters. Calibration of the bivariate model can then be assessed by examining the marginal uniformity of $U_1$ and $U_2$. Empirical deviations from uniformity provide evidence of model misspecification.

In practice, however, such multivariate PIT diagnostics have rarely been applied in the presence of censoring, as censoring complicates the transformation. When the event time $T$ is subject to right-censoring, the PIT value $F_T(T)$ for a censored observation is only known to lie within the interval $[F_T(T), 1]$. To address this, we use a Kaplan–Meier (KM) estimator, where PIT values for censored observations are treated as random draws from $U(F_T(T), 1)$~\citep{klugman1999fitting}. In the presence of covariates, this corresponds to using the KM estimate of $F_T(T \mid \rX)$.

For the conditional PIT value $F_{Y \mid T, \rX}(Y \mid T > \underline{t}, \rX)$, the complication again arises due to censoring of the event time. When the observed data consist of $O = (y, \underline{t}, \rx)$, indicating that the event time is right-censored at $\underline{t}$, we estimate the conditional distribution of $Y$ given $T > \underline{t}$ and covariates $\rX = \rx$ using
\begin{align*}
	P(Y \leq y \mid T > \underline{t}, \rX = \rx) = \frac{F_{Y \mid \rX}(y \mid \rx) - F_{Y, T \mid \rX}(y, \underline{t} \mid \rx)}{1 - F_{T \mid \rX}(\underline{t} \mid \rx)}.
\end{align*}
This expression enables the computation of PIT values for censored event times while appropriately conditioning on covariates. These PIT-based tools offer an approach for assessing model calibration, both in the presence and absence of censoring, and under covariate adjustment. We demonstrate their use in the application that follows in Section~\ref{sec:application}. 
\section{Empirical evaluation}
\label{sec:simulation}

In this section, we use simulated data to evaluate the performance of our NPB model in assessing the prognostic accuracy of biomarkers. Since most existing methods for time-dependent ROC analysis do not account for covariates, we begin by comparing these methods in the unconditional scenario, followed by a comparison when covariates are introduced. For competitor methods in the unconditional case, we considered actively maintained \proglang{R} packages that are
hosted and systematically checked on the Comprehensive \proglang{R} Archive Network (CRAN), as summarized in Table~\ref{tab:roc_methods}. For a review of these methods, see \AYcite{Kamarudin et al.}{kamarudin2017time}.

\begin{table}[h]
	\centering
	\begin{tabular}{lp{6cm} p{4.5cm}}
		\toprule
		{Package}       & {Methods} & {Reference} \\
		\midrule
		\pkg{survivalROC}   & Nearest neighbor estimation (NNE), Kaplan-Meier (KM) & \AYcite{Heagerty et al.}{heagerty2000time}\\
		\pkg{timeROC}       & Inverse probability weighting (IPW) & \AYcite{Blanche et al.}{blanche2013estimating} \\ 
		\pkg{smoothROCtime} & Bivariate kernel density estimation & \AYcite{Mart\'{\i}nez-Camblor and Pardo-Fern\'andez}{martinez2018smooth}\\ 
		\pkg{nsROC}         & Cox proportional hazards model (COX), Kaplan-Meier (KM) and weighted Kaplan-Meier (WKM) & \AYcite{P\'erez-Fern\'andez et al.}{perez2018nsroc} \\
		\pkg{cenROC}        & Weighted kernel smoother with (TR) and without boundary correction (UTR) & \AYcite{Beyene and El Ghouch}{beyene2020smoothed} \\ 
		\pkg{tdROC}         & Nonparametric weight adjustments & \AYcite{Li et al.}{li2018simple}\\ 
		\bottomrule
	\end{tabular}
	\caption{\proglang{R} packages and methods for unconditional time-dependent ROC analysis.}
	\label{tab:roc_methods}
\end{table}
We label each method using the \proglang{R} package name, followed by an abbreviation for the specific estimator when multiple options are available within the same package. 

\subsection{Data generating process}

For the biomarker, we as data generating process (DGP) considered the following distributions $F_Y$: standard
normal $N(0, 1)$, two-component normal mixture of $N(1, 1)$ and $N(4,
1.5^2)$ with equal mixture weights, and the Chi-squared with three degrees
of freedom.  For event time, we used the following distributions $F_T$:
standard lognormal, Weibull with shape $1.4$ and rate $2.0$, and Gamma with
shape $1.5$ and rate $1.2$.  We explored all combinations of biomarker and event
time distributions, with sample sizes of $N \in \{100, 500, 1000\}$ for a
$1000$ replications.  This data generation approach is similar to that used in
previous studies \citep{martinez2018smooth, yu2019optimal, beyene2024time}.

We generated $\mZ = (Z_1, Z_2)^\top$ from a standard bivariate normal distribution with correlation coefficient $\rho \in \{-0.3, -0.5\}$. The negative correlation implies that higher baseline biomarker concentrations are associated with shorter event times. To explore the range of marginal distributions, we transformed the marginals to uniform distributions $(U_1, U_2) = \left(\Phi(Z_1), \Phi(Z_2)\right)$ and applied different quantile functions to obtain the desired distributions of the biomarker and event time $(Y, T)$. This preserves the dependence structure governed by $\rho$ while allowing the marginal distributions to vary.


To achieve an expected censoring rate of 50\%, the censoring time $C$ can be generated independently following the same distribution as the event time. For any other censoring rate $\kappa = \Prob(T > C)$, independent realizations can be drawn from a distribution of the same form as the event time, given by $F_C(c) = \Phi\left( \hT(c) + a \right)$, where $a = \Phi^{-1}(\kappa) \sqrt{2}$ is an offset based on the definition of the AUC used to achieve the target censoring rate \citep{sewak2023estimating}.  In our study, we applied censoring rates of $\kappa \in \{0.3, 0.5\}$.

\subsection{Unconditional results}

\begin{figure}[t!]
	\centering
	\includegraphics[width=0.85\linewidth]{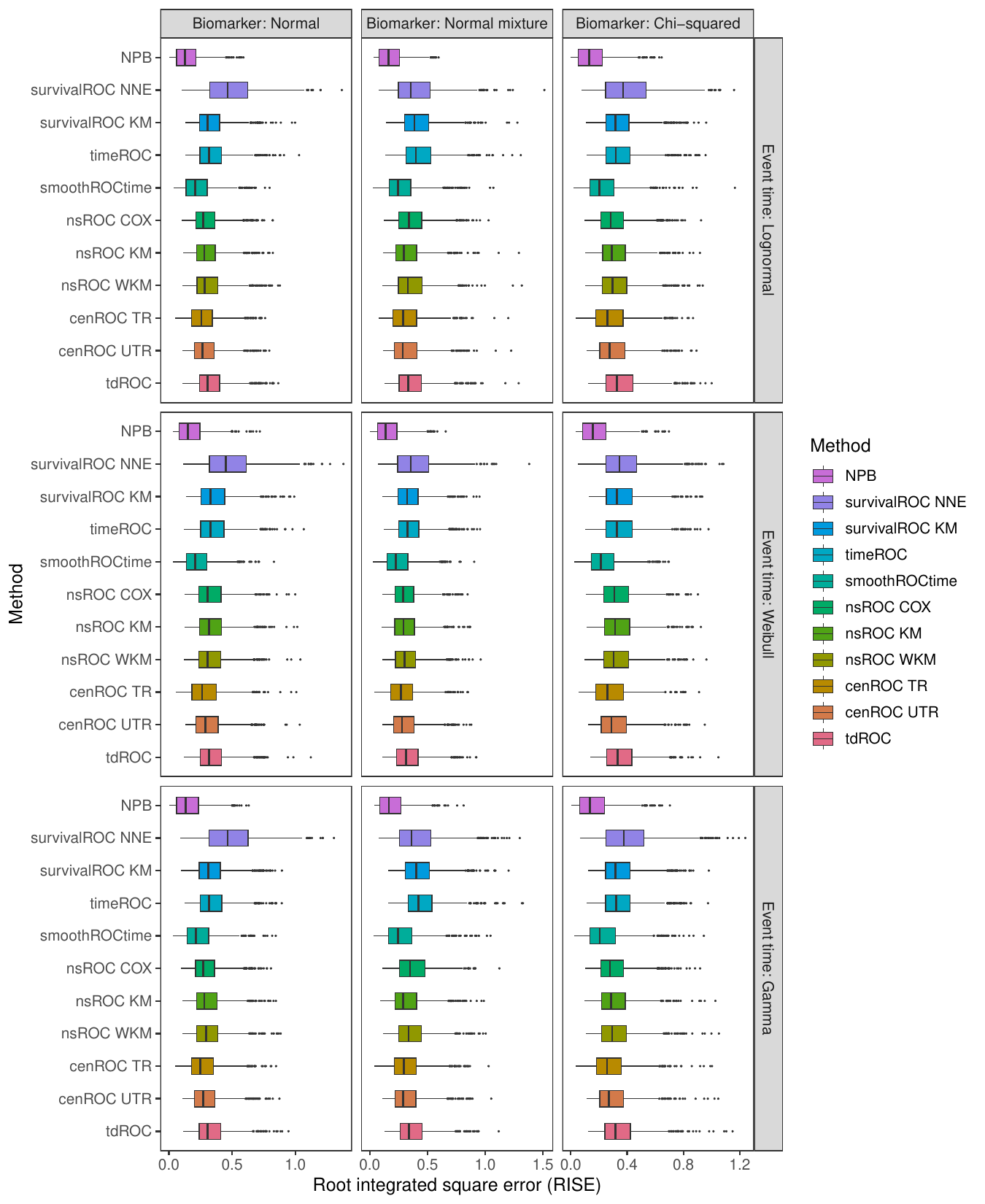} 
	\caption{Distribution of root integrated squared errors (RISE) for unconditional ROC estimators at the median time quantile with a sample size of $N = 500$, event time distributions of $\{\text{Lognormal}, \text{Weibull}, \text{Gamma}\}$, 30\% censoring rate and a correlation of $\rho = -0.5$ between transformed biomarker and event time distributions.}
	\label{fig:sim_roct}
\end{figure}

We evaluated the AUC at times corresponding to the unconditional quantiles
$0.1$, $0.25$, $0.5$, and $0.75$ for each estimator and biomarker
distribution.  Figure~\ref{fig:sim_roct} displays the mean and standard
error across the $1000$ repetitions of the estimated AUC for different methods
and sample sizes.  The NPB method shows a small bias for small sample sizes
but is generally unbiased for larger sample sizes across different time
quantiles and biomarker distributions.  Additionally, the NPB model
demonstrates lower variability compared to other methods.  In contrast, the
NNE estimator from \pkg{survivalROC} exhibits bias in all scenarios.  Most
methods are unbiased for time quantiles below $0.5$, but \pkg{smoothROCtime}
and \pkg{nsROC} have some bias for higher time quantiles.  The results were
consistent across different event time distributions.  Results for a
lognormal event time distribution, a $50\%$ censoring rate and a correlation
of $\rho = -0.7$ are shown in Supplementary Figure~\ref{fig:app_sim_auct}.

\begin{figure}[t!]
	\centering
	\includegraphics[width=0.85\linewidth]{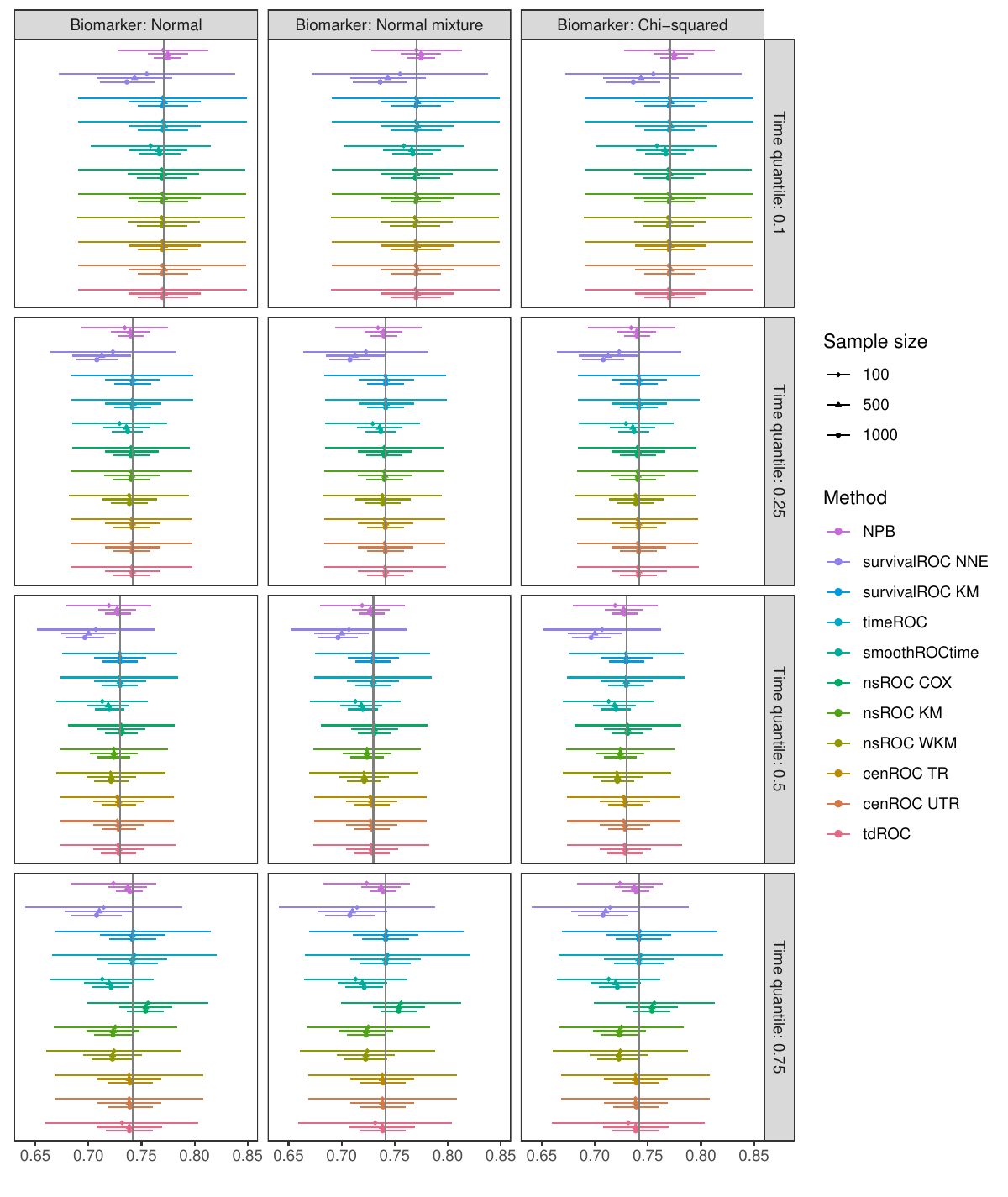} 
	\caption{Mean and standard deviation of the unconditional AUC for each method with sample sizes of $N = \{100, 500, 1000\}$, Weibull event time distribution, 30\% censoring rate and a correlation of $\rho = -0.5$ between transformed biomarker and event time distributions. The evaluation is across different biomarker distributions (Normal, Normal mixture, Chi-squared) and time quantiles
	$(0.1, 0.25, 0.5, 0.75)$. }
	\label{fig:sim_auct}
\end{figure}

For the ROC curve, we calculated the root integrated squared error (RISE)
\begin{align*}
	\text{RISE} = \sqrt{\int_0^1 \left( \widehat{\ROC}_t(p) - \ROC_t(p) \right)^2 \, d p}
\end{align*}
where  $\widehat{\ROC}_t$ is the estimated ROC curve and $\ROC_t$ is the
true curve for a given time point~$t$.  Figure~\ref{fig:sim_roct} displays
the RISE distribution for the time quantile $0.5$ for different biomarker and
event time distributions.  The NPB method achieves lowest RISE values
across all setups, indicating better performance in estimating the ROC
curve.  Methods like \pkg{survivalROC} (NNE and KM) and \pkg{timeROC} show
higher RISE values, suggesting less precise estimates.  Kernel-based methods
such as \pkg{cenROC} and \pkg{smoothROCtime} also perform well, but do not
outperform NPB, particularly for more complex biomarker distributions like
the normal mixture and chi-squared.  Supplementary Figure~\ref{fig:app_sim_roct} 
provides additional results for different sample sizes with a $50\%$ censoring
rate and a correlation of $\rho=-0.7$.

\subsection{Conditional results}

For the conditional case, we generated a continuous covariate from a standard uniform distribution $X \sim U(0, 1)$. To ensure consistency with the marginal distributions from the previously described unconditional data generating process and to incorporate the covariate shift on a standardized scale, we adjusted $\mZ$ to $(Z_1 + \gamma_Y X, Z_2 + \gamma_T X)$ where $\gamma_Y = 0.5$ and $\gamma_T = 3$. We then applied the standard normal CDF followed by the appropriate quantile function of the marginal distribution.

We compare our NPB approach to the only existing method for time-dependent ROC analysis with covariates that has available code. The \proglang{R} package \pkg{CondTimeROC} implements a nonparametric kernel smoothing technique for one continuous covariate \citep{rodriguez2016nonparametric}. There are several estimators implemented in the package: smooth with bandwidth selected using a plug-in approach \citep{altman1995bandwidth} or data-driven \citep{li2013optimal} and completely nonparametric. The package also implements the semiparametric technique of \AYcite{Song and Zhou}{song2008semiparametric}.

\begin{figure}[t!]
	\centering
	\includegraphics[width=0.85\linewidth]{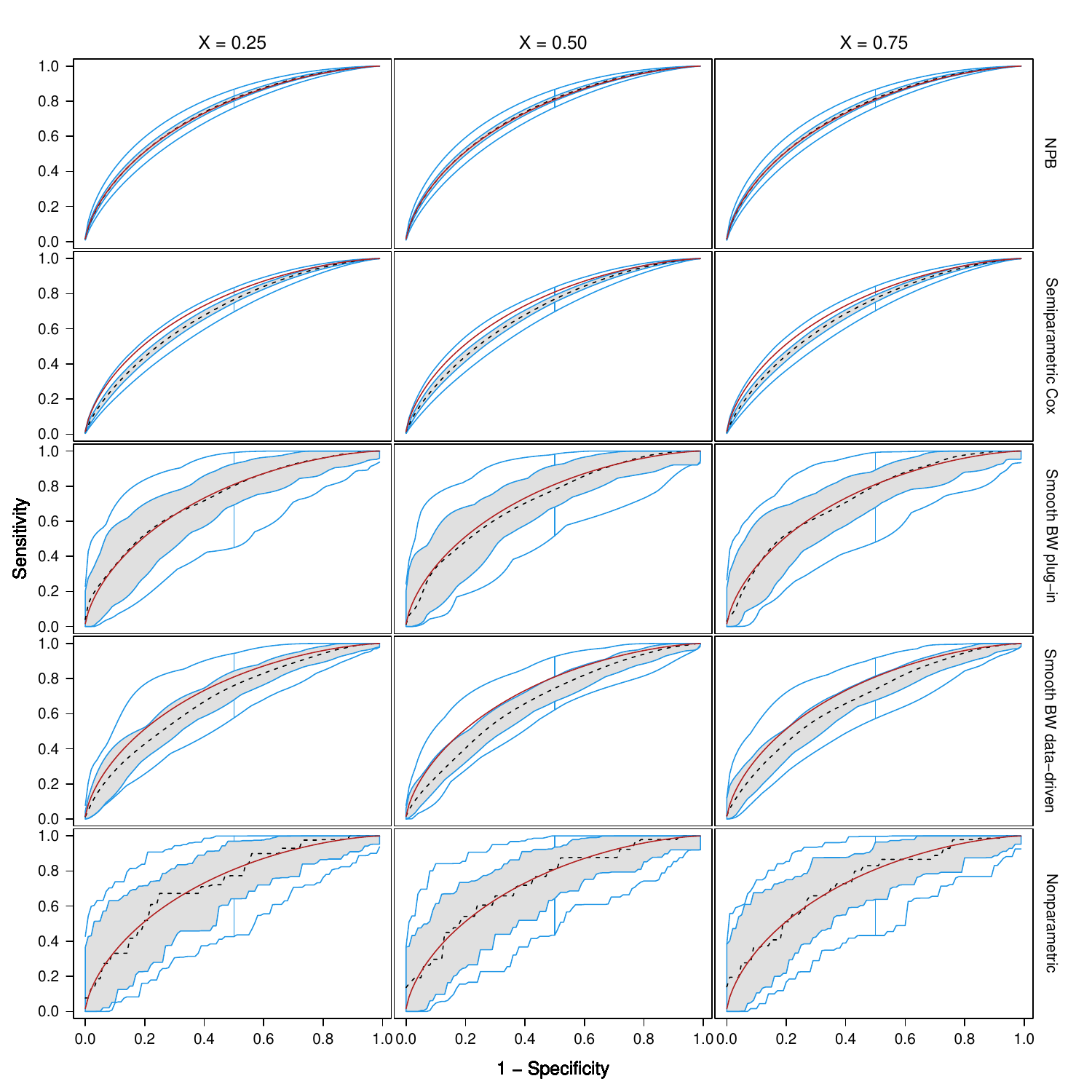} 
	\caption{Functional boxplots for covariate-specific time dependent ROC curves at the median time quantile, with a sample size of $N = 500$, Weibull event time distribution, 30\% censoring rate, and a correlation of $\rho = -0.5$ between transformed biomarker and event time distributions. The red line indicates the true covariate-specific ROC curve for each case.}
	\label{fig:sim_roctx}
\end{figure}

For the various methods, we evaluated the covariate-specific ROC curve
$\ROC_t(x)$ at different values of the covariate using functional boxplots,
shown in Figure~\ref{fig:sim_roctx}.  These plots summarize the ROC
functions over $1000$ replications.  The dashed line represents the median
function, the grey area highlights the $50\%$ central region and the blue
curves are analogous to whiskers in a traditional boxplot.  The red curve
represents the ROC curve from the true data-generating process.  To ensure
comparable sensitivity and specificity calculations across methods, we
varied $t$ based on covariates so that each analysis included an equal
number of uncensored observations.  This adjustment was necessary for the
nonparametric methods, which had high variance when only a few data points
were available for estimation.  The NPB model estimates are unbiased and
show low variance.  In contrast, the ROC estimates from the smoothing method
appear sensitive to the choice of bandwidth, leading to bias in some cases. 
The nonparametric estimates are unbiased but show high variability, while
the semiparametric estimates are biased, likely due to the violation of the
proportional hazards assumption.

We also evaluated the performance of parameter estimates from our NPB model
in the presence of multiple covariates.  For this, we generated two
continuous covariates from a standard uniform distribution and one binary
covariate.  Supplementary Figure~\ref{fig:sim_auctx} presents the bias of the
parameter estimates under different sample sizes and censoring proportions. 
The likelihood-based inference procedures produced unbiased and
efficient estimates.  However, it is important to note that when the
dependency structure changes or when the covariates include nonlinear terms,
the estimates can become biased. The performance of the NPB under model
misspecification is evaluated in Supplementary Material \ref{sec:misspec}.

\section{Application to prognostic biomarkers in ALS}
\label{sec:application}
In this section, we apply our framework to data from our motivational study on prognostic biomarkers for ALS \citep{benatar2020validation}. Serum NfL values in the current report were corrected for a 4-fold dilution factor that was inadvertently omitted in the previous publication.
We apply the NPB model to determine the prognostic accuracy of baseline serum NfL concentration in predicting survival time in ALS. Our analysis begins with an unconditional model and we then incorporate baseline covariates to assess their impact on prognostic accuracy.

\subsection{Unconditional NPB model}

We estimated the unconditional NPB model to evaluate the prognostic accuracy of baseline neurofilament light (NfL) concentration for predicting survival time in patients with ALS. Supplementary Figure~\ref{fig:densu} displays the estimated joint density of the biomarker and event time, showing good agreement with the observed data. The model successfully captures the negative association between baseline NfL levels and survival time. Model diagnostics using QQ plots of the PIT values (Supplementary Figure~\ref{fig:pitu}) revealed no evidence of major misspecification.

Using the fitted model, we derived the time-dependent ROC curve and associated summary measures, including the area under the curve (AUC), Youden index, optimal threshold, sensitivity and specificity. The optimal NfL threshold corresponds to the point at which the sum of sensitivity and specificity is maximized. Table~\ref{tab:sumu} presents the estimated summary statistics along with their corresponding 95\% confidence intervals. The results indicate that the prognostic utility of baseline NfL concentration decreases over longer prediction horizons

\begin{table}[ht] \centering
	\resizebox{\columnwidth}{!}{
		\begin{tabular}{llllll}
			\toprule
			Time & AUC & Youden index& Optimal threshold & Sensitivity & Specificity \\ 
			(months) & (95\% CI) & (95\% CI)  & (95\% CI) & (95\% CI) & (95\% CI) \\ 
			\midrule
			3 &  0.78 (0.70, 0.86) &  0.42 (0.29, 0.55) & 92.93 ( 77.82, 108.04) &  0.73 (0.66, 0.80) &  0.69 (0.63, 0.76) \\ 
			6 &  0.75 (0.68, 0.83) &  0.37 (0.26, 0.49) & 84.85 ( 72.50,  97.19) &  0.71 (0.65, 0.78) &  0.66 (0.61, 0.71) \\ 
			12 &  0.71 (0.65, 0.78) &  0.31 (0.21, 0.41) & 76.77 ( 65.71,  87.83) &  0.66 (0.60, 0.72) &  0.65 (0.60, 0.70) \\ 
			24 &  0.70 (0.63, 0.76) &  0.28 (0.19, 0.38) & 64.65 ( 55.38,  73.92) &  0.65 (0.60, 0.70) &  0.63 (0.58, 0.68) \\ 
			\bottomrule
		\end{tabular}
	}
	\caption{Model-based summary statistics describing the prognostic accuracy of baseline serum NfL concentration for predicting survival at different time points. The table includes the AUC, Youden index, optimal threshold for NfL concentration, sensitivity and specificity at the optimal threshold, each presented with their corresponding 95\% confidence intervals (CI) at
	$3$, $6$, $12$, and $24$ months.}
	\label{tab:sumu}
\end{table}

\begin{figure}[t!]
	\centering
	\includegraphics[width=\linewidth]{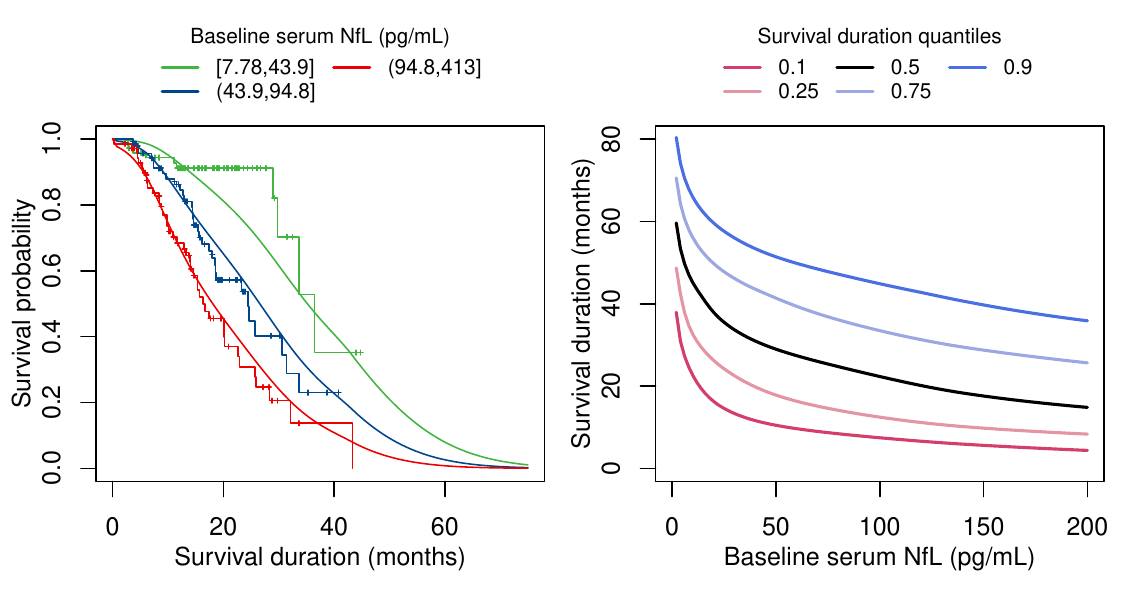} 
	\caption{Left: Survival curves stratified by tertiles of baseline serum neurofilament light (NfL) concentration. Right: Survival time quantiles as a function of baseline serum NfL concentration.}
	\label{fig:surv_qns}
\end{figure}

Access to the joint distribution allows us to estimate the survival time distribution for specific ranges of baseline NfL concentrations. NfL levels were categorized into tertiles (low, medium, high), as shown in Figure~\ref{fig:surv_qns}. The results depict an association between higher NfL levels and shorter survival times, with the highest tertile exhibiting the steepest decline in survival probability. The model also provides estimates for median survival time and other quantiles, showing that higher NfL levels predict shorter survival times across all quantiles. Additionally, the rate of decline in the post-ALSFRS-R score is strongly associated with ALS survival. Individuals with higher NfL levels exhibit a more rapid decline in ALSFRS-R scores, as shown in 
Supplementary Figure~\ref{fig:longit}.


\subsection{Conditional NPB model}
While the unconditional model provides an overall view of the relationship between NfL and survival, incorporating covariates allows us to account for confounding effects and assess how these factors influence the prognostic accuracy of NfL. We considered several baseline factors that could influence the prognostic accuracy of NfL for ALS.

The mean baseline age of patients was $60.1$ years (SD $11.7$).  The cohort comprised $93$ females ($42.7\%$) and $125$ males
($57.3\%$). 
In terms of genetic characteristics, the C9orf72 repeat expansion was
present in $24$ participants ($11.0\%$) and absent in $194$ patients
($89.0\%$). 
Site of symptom onset was bulbar only for $42$ participants ($19.3\%$), limb only for $154$ ($70.6\%$), and other regions or multi-region for $22$ ($10.0\%$).
The median baseline ALS Functional Rating Scale -- Revised (ALSFRS-R) score was $36.0$ points (IQR $9.0$). We also assessed the past (i.e. pre-baseline) rate of disease progression, defined by
\begin{align*}
	\Delta\text{FRS} = \frac{48 - \text{ALSFRS-R}_\text{baseline}}{\text{Months from onset to baseline}},
\end{align*}
where 48 is the maximum score on the ALSFRS-R. 
The average $\Delta\text{FRS}$ was $0.60$ points per month (SD $0.47$).

For ease of the following visualizations, we fixed continuous variables to
their mean values and categorical variables to their modal values, unless
otherwise specified.  We fit the conditional NPB model with the marginal
model for NfL specified as
\begin{align*}
	h_Y(Y_i \mid \parm) &= \beta_{Y,1} \text{Age}_i +f \beta_{Y,2} \text{Sex}_i + \beta_{Y,3} \text{C9}_i + \beta_{Y,4} \Delta \text{FRS}_i + \beta_{Y,5} \text{ALSF}_i + \beta_{Y,6} \text{Site}_i + Z
\end{align*}
where the baseline transformation function is given by $h_Y(y \mid \parm) = \sum_{m=0}^6 \vartheta_{Y,m}b_m(\log(y))$ and $Z \sim N(0, 1)$. The survival time model includes the same set of covariates and the marginal transformation function for the survival time follows an analogous Bernstein polynomial form. 

Supplementary Figure~\ref{fig:trafo} displays the estimated baseline transformation functions for both NfL and survival time. These are nonlinear highlighting the need of having transformation functions for both margins. Supplementary Figure~\ref{fig:pitc} displays the QQ plots of PIT values. These indicate a generally good fit, though there may be slight departures in the upper tail of the survival distribution, likely due to reduced data support at longer follow-up times.

The coefficients of the joint model are presented in Table~\ref{tab:coefs}. 
The unconditional correlation indicates whether there is a significant
association between the biomarker and event time.  If the confidence
interval for this correlation includes zero, the biomarker likely lacks
prognostic accuracy.  For NfL, the unconditional correlation coefficient is
relatively strong at $-0.429$, with a $95\%$ confidence interval of $-0.283$
to $-0.547$, indicating a significant association with survival duration. 
The conditional correlation shows a reduction compared to the unconditional
correlation, suggesting that some of the covariates account for a portion of
the variation in survival duration.

\begin{table}[ht]
	\centering
	\begin{tabular}{p{5cm}cc}
		\toprule 
		\multirow{2}{*}{Variable} & {NfL} & {Survival time} \\
		\cmidrule(lr){2-3} & \multicolumn{2}{c}{Coefficient (95\% CI)} \\
		\midrule 
		Unconditional correlation & \multicolumn{2}{c}{ $-0.428 \, (-0.283, -0.547)$ } \\
		Conditional correlation & \multicolumn{2}{c}{ $-0.324 \, (-0.149, -0.471)$ } \\
		{Covariates} $\beta_i$ & & \\
		\quad Age & $\phantom{+}0.020 \, (\phantom{+}0.010, \phantom{+}0.031)$ & $-0.021 \, (-0.036, -0.006)$ \\ 
		\quad Sex - Male & $\phantom{+}0.565 \, (\phantom{+}0.290, \phantom{+}0.840)$ & $\phantom{+}0.142 \, (-0.239, \phantom{+}0.522)$ \\ 
		\quad C9orf72 - Positive & $\phantom{+}0.389 \, (-0.055, \phantom{+}0.833)$ & $-0.541 \, (-1.076, -0.006)$ \\ 
		\quad $\Delta$FRS & $\phantom{+}0.801 \, (\phantom{+}0.467, \phantom{+}1.135)$ & $-0.513 \, (-0.885, -0.142)$ \\ 
		\quad Baseline ALSFRS-R & $\phantom{+}0.000 \, (-0.021, \phantom{+}0.021)$ & $\phantom{+}0.035 \, (\phantom{+}0.015, \phantom{+}0.054)$ \\ 
		\quad Site - Bulbar only & $\phantom{+}0.700 \, (\phantom{+}0.284, \phantom{+}1.116)$ & $-0.282 \, (-0.721, \phantom{+}0.157)$ \\ 
		\quad Site - Other & $\phantom{+}0.284 \, (-0.129, \phantom{+}0.697)$ & $\phantom{+}0.649 \, (-0.067, \phantom{+}1.366)$ \\ 
		\bottomrule
	\end{tabular}
	\caption{Estimated coefficients of the nonparanormal prognostic biomarker model along with their corresponding 95\% confidence intervals (CI).}
	\label{tab:coefs}
\end{table}

\begin{figure}[t!]
	\centering
	\includegraphics[width=0.9\linewidth]{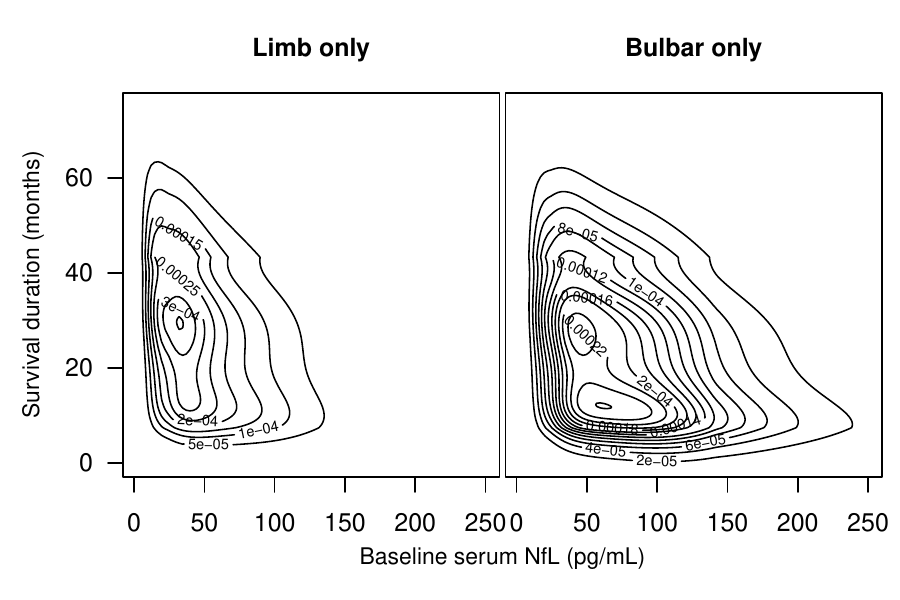} 
	\caption{Bivariate density of baseline serum neurofilament light (NfL) concentration and survival time by the presence of the site of onset (limb only vs. bulbar only). The plot is based fixed covariate values for a 60-year-old male with a negative C9orf72 status, a progression rate ($\Delta$FRS) of
	$0.60$ points/month and a baseline ALSFRS-R score of $35$.}
	\label{fig:dens}
\end{figure}
Figure~\ref{fig:dens} shows the estimated joint distribution of baseline NfL
concentration and survival time by the site of ALS onset.  In both groups,
higher baseline NfL concentrations are generally associated with shorter
survival times.  Patients with bulbar-onset ALS show lower survival times
and higher NfL concentrations compared to those with limb-onset ALS.  This
aligns with previous research, which suggests that bulbar-onset ALS
typically has a more aggressive disease
progression~\citep{zoccolella2008analysis}.

\begin{figure}[t!]
	\centering
	\includegraphics[width=0.9\linewidth]{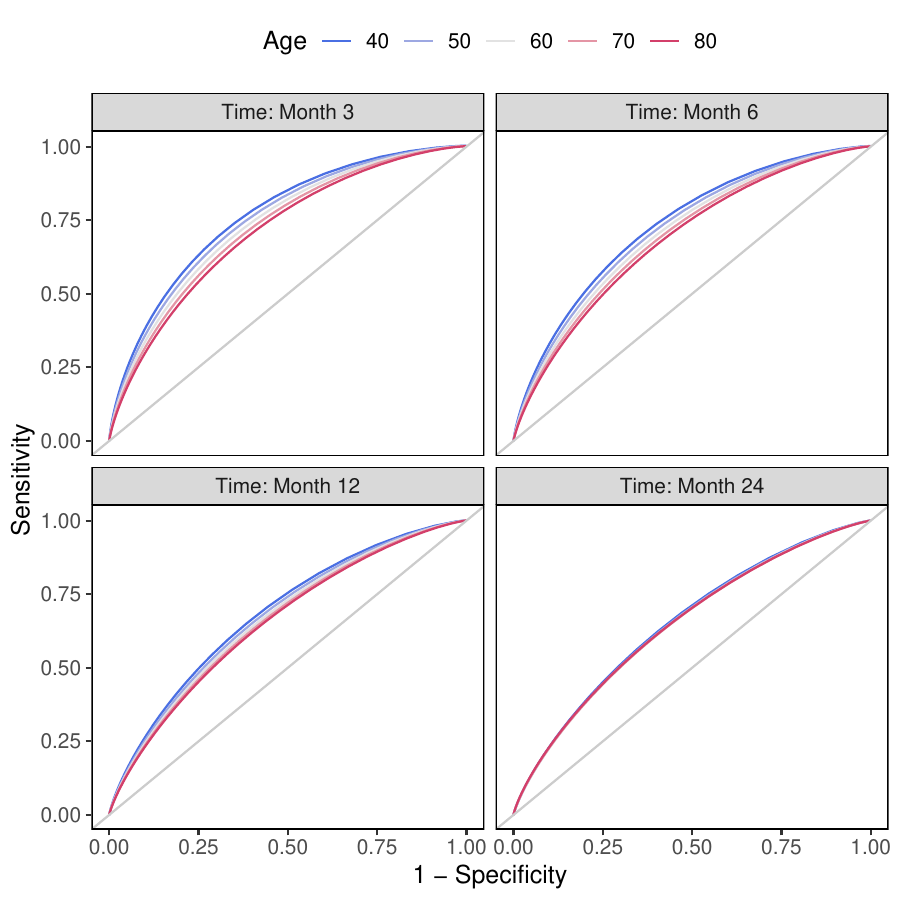} 
	\caption{Covariate-specific time-dependent ROC curves by age and the presence of the C9orf72 repeat expansion for a female with a $\Delta$FRS of
	$0.5$. The plot is based fixed covariate values for a male with a negative C9orf72 status, a progression rate ($\Delta$FRS) of
	$0.60$ points/month and a baseline ALSFRS-R score of $35$.}
	\label{fig:troc}
\end{figure}

Figure~\ref{fig:troc} shows the estimated age-specific time-dependent ROC
curves.  The accuracy of baseline NfL in predicting survival declines as the
time horizon increases.  For any given time horizon, the prognostic accuracy
of NfL decreases with increasing age.  Further, age appears to have a
greater impact when predicting early ALS prognosis compared to later stages
of the disease.

\begin{figure}[t!]
	\centering
	\includegraphics[width=0.9\linewidth]{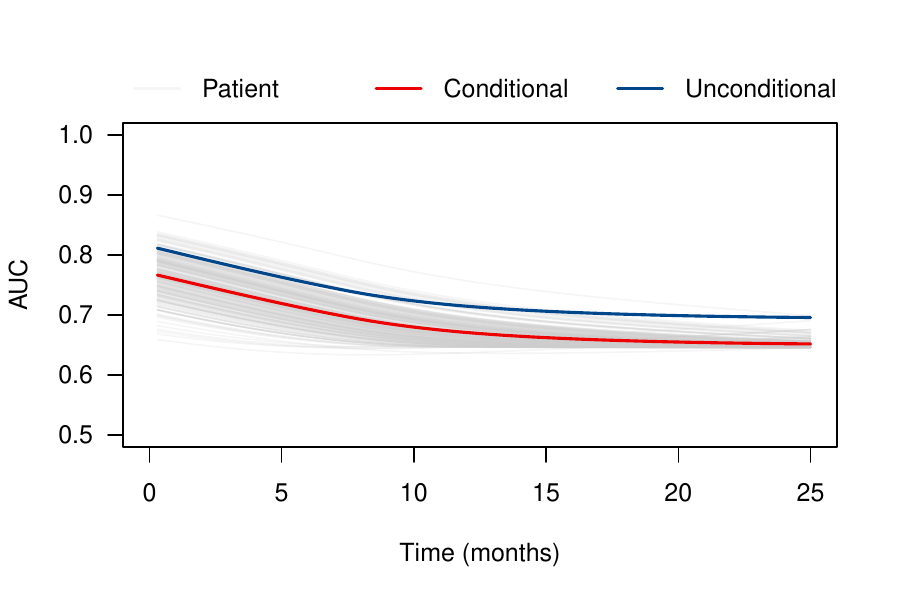} 
	\caption{Spaghetti plot for patient-specific time-dependent AUCs (in gray), smoothed average across all patients (red) and unconditional (blue).}
	\label{fig:marg_auc}
\end{figure}

The effect of different covariate combinations on prognostic accuracy
varies, making it challenging to summarize these effects with a single
model-based statistic or covariate-specific ROC curves. 
Figure~\ref{fig:marg_auc} shows a spaghetti plot, where each line represents
the time-dependent AUC for an individual patient’s unique combination of
covariates.  The red line shows the smoothed average across all patients,
while the blue line corresponds to the AUC from the unconditional model. 
Even after adjusting for covariates, the prognostic accuracy of baseline NfL
declines over longer time horizons.  However, incorporating covariates
reduces the prognostic accuracy of NfL.  Failing to account for these
factors may lead to overly optimistic estimates.

Taken together, our findings emphasize the importance of covariate adjustment when evaluating prognostic biomarkers. The conditional NPB model revealed that the prognostic value of NfL is lower among patients with poorer prognostic factors such as older age, C9orf72-positive status, and higher $\Delta$FRS. In contrast, factors such as sex, site of onset, and baseline ALSFRS-R showed limited influence on prognostic accuracy.


\section{Discussion}
\label{sec:conclusion}

In this paper, we present a flexible statistical framework for evaluating the prognostic accuracy of biomarkers over time, while accounting for patient characteristics. We construct time-dependent ROC curves and summary indices using model-based covariate-specific sensitivity and specificity. The proposed NPB model accommodates a broad class of functional forms for the marginals and handles both right- and interval-censoring within a unified likelihood-based approach.
Simulation studies demonstrate that the method performs well even in small to moderate sample sizes. Model calibration can be evaluated using the proposed PIT-based diagnostics, including QQ plots adapted to censored data and covariate-specific distributions.

%

One of the key advantages of our approach is its capacity to incorporate multiple covariates while retaining statistical efficiency. The parametric transformation models for the marginals yield near semiparametric-efficient estimators even in small samples~\citep{hothorn2024on}, which translates to low-variance estimates of time-dependent AUC and other ROC-based measures, as shown in our simulations. By using a full-likelihood formulation rather than a two-stage or surrogate approach, we obtain closed-form score functions and enable valid likelihood-based inference, including confidence for all model parameters. The same framework also handles a wide range of data complexities, including detection limits, mixed outcome types and various forms of censoring without the need for combing different estimators to handle each setting individually.

Several areas warrant further investigation. As with any joint modeling framework, the validity of the results depends on the adequacy of the model assumptions. Our simulations show that while the NPB model performs well under correlation-driven dependence structures, it can yield biased results when the relationship between biomarker and event time is a direct functional form (Supplementary Figure~\ref{fig:sim_roctx_ms}). To address this, we proposed a PIT-based diagnostic approach to assess the model calibration. These diagnostics are adapted for censored data and covariates, and showed no substantial misspecification in our ALS application. Also, although we focused on the Gaussian copula in the paper, the framework can be extended to alternative copula families and future work should explore such generalizations.

We chose a linear predictor for covariates to maintain simplicity. However, the framework is flexible and can be extended to accommodate non-linear effects through modifications to the marginal models. Additionally, while our method supports proportional hazards when appropriate, it is not limited to this assumption and allows for broader modeling choices. Covariate selection remains an important area for further development. Although model selection based on out-of-sample log-likelihoods is theoretically possible, it can be computationally demanding and impractical in smaller datasets. Rather than serving as a fully automated selection procedure, our framework is intended to support informed modeling decisions guided by clinical knowledge.

In practice, the choice of method for time-dependent ROC analysis involves a bias-variance trade-off. Fully nonparametric approaches are asymptotically unbiased and flexible but often suffer from high variance in small samples and typically cannot handle more than one covariate. In contrast, the proposed likelihood-based approach offers efficient estimation and covariate adjustment, but relies on model assumptions. Hybrid methods that strike a balance between flexibility and efficiency present an important area for future work.

We do not suggest our method as a default. In large datasets with minimal censoring and low-dimensional covariates, nonparametric or kernel-based approaches may suffice and require fewer assumptions. However, our method is particularly valuable in scenarios with limited sample sizes, complex censoring and covariate-driven heterogeneity in biomarker performance or when formal inference on model parameters is required.

As the number of candidate biomarkers continues to grow, so too does the need for robust statistical tools to evaluate their prognostic utility. The NPB framework provides a flexible and efficient solution for quantifying how biomarker performance varies across patient subgroups. This can support better risk stratification and can ultimately lead to more efficient clinical trials, by enabling more targeted inclusion criteria and reducing required sample sizes.





\section*{Computational details}

We used \pkg{mvtnorm} \citep[][version
1.3.4]{pkg:mvtnorm} to sample from multivariate
normal distributions, \pkg{pracma} \citep[][version
2.4.4]{pkg:pracma} for numerical integration of ROC
curves, and \pkg{qrng} \citep[][version
0.0.10]{pkg:qrng} for random number generation.

To evaluate competitor methods in our simulation studies, we used the following \proglang{R} packages: \pkg{survivalROC}~\citep[][version 1.0.3.1]{pkg:survivalROC} for nearest-neighbor and Kaplan-Meier estimations;
\pkg{timeROC}~\citep[][version 0.4]{blanche2013estimating} for inverse probability weighting;
\pkg{smoothROCtime}~\citep[][version 0.1.0]{pkg:smoothROCtime} for bivariate kernel density estimation;
\pkg{nsROC}~\citep[][version 1.1]{pkg:nsROC} for proportional hazards-based methods;
\pkg{cenROC}~\citep[][version 2.0.0]{pkg:cenROC} for smoothed kernel methods;
\pkg{tdROC}~\citep[][version 2.0]{pkg:tdROC} for nonparametric kernel smoothing weights.
Each package was used to implement its respective method for estimating time-dependent ROC curves.

All computations were performed using \textsf{R} version
4.5.1
\citep{R}.

A reference implementation of the nonparanormal prognostic biomarker model is 
available in the \pkg{tram} add-on package \citep{pkg:tram} to the \proglang{R} system for
statistical computing. The simulation results presented in Sections~\ref{sec:simulation}
are available on \url{https://gitlab.com/asewak/npb}. The application to 
prognostic biomarkers in ALS in Section~\ref{sec:application} can be reproduced by
running the code from within \proglang{R}:
\begin{knitrout}
\definecolor{shadecolor}{rgb}{0.969, 0.969, 0.969}\color{fgcolor}\begin{kframe}
\begin{alltt}
\hlkwd{install.packages}\hldef{(}\hlsng{"tram"}\hldef{)}
\hlkwd{library}\hldef{(}\hlsng{"tram"}\hldef{)}
\hlkwd{demo}\hldef{(}\hlsng{"npb"}\hldef{,} \hlkwc{package} \hldef{=} \hlsng{"tram"}\hldef{)}
\end{alltt}
\end{kframe}
\end{knitrout}

\section*{Acknowledgements}

Torsten Hothorn received financial support from the Swiss National
Science Foundation (grant number  200021\_219384). 
\clearpage


\bibliography{refs, packages}

\clearpage


\begin{appendix}
\section{Alternative ROC curve definitions}
\label{app:alt_roc}

The NPB framework introduces an approach to time-dependent ROC analysis by modeling the joint distribution of the biomarker and event time conditional on covariates. This strategy allows for derivation of various ROC curve types, including incident-dynamic and incident-static, as well as associated summary indices such as the AUC and Youden Index. Unlike many existing methods, which are tailored to specific definitions of sensitivity or specificity, the NPB framework provides a way to derive all forms of time-dependent ROC curves.

\subsection{Incident-dynamic}
The \emph{incident} sensitivity is the probability that a subject’s biomarker value exceeds a threshold $c$ at the exact time they experience the event 
\begin{align*}
	\Sei_t(c \mid \rx) &= \Prob(Y > c \mid T = t, \rX = \rx)
\end{align*}
In this definition, cases are defined as subjects experiencing the event at time $t$, while subjects who experienced the event before $t$ are excluded from the calculation. This leads to the \emph{incident-dynamic} ROC curve, which evaluates the biomarker's ability to distinguish between subjects experiencing the event at $T = t$ and those who have not yet experienced it. This measure is particularly valuable for guiding treatment decisions at multiple time points.

Under the NPB framework, where the conditional distribution of the biomarker given event time is normal, incident sensitivity has a closed-form expression
\begin{align*}
	\Sei_t(c \mid \rx) &= \Phi \left( \frac{\rho \hT(t \mid \rx) - \hY(c \mid \rx)}{\sqrt{1 - \rho^2}}\right) \\
	&= \Phi \left( \frac{ \rx^\top\betavec_Y + \rho \left(\hT(t) - \rx^\top \betavec_T \right) - \hY(c)}{\sqrt{1 - \rho^2}}\right).
\end{align*}
The second form assumes linear transformation models for the marginal distributions of the biomarker and event time, as introduced in Section~\ref{sec:npbm}.
Alternative marginal models can also be incorporated as discussed in Section~\ref{sec:extension}.
Note that in this setup incident sensitivity involves only the evaluation of a univariate distribution function.

\subsection{Incident-static}
\emph{Static} sensitivity is particularly useful for comparing incident cases to control subjects who are long-term survivors, i.e., those who remain event-free throughout a fixed follow-up period $(0, t^*)$~\citep{etzioni1999incorporating}. It is defined as
\begin{align*}
	\Sps(c \mid \rx) &= \Prob(Y \leq c \mid T > t^*, \rX = \rx)
\end{align*}
which leads to the \emph{incident-static} ROC curve.
Under the NPB framework, the joint model for the biomarker and event time remains the same as for the dynamic specificity case described in the main text. The key difference lies in the fact that $\Sps$ is no longer dependent on the value of $t$. Instead it focuses on a fixed follow-up time $t^*$ which is prespecified to a value which is considered a long enough time to observe the event of interest.
The dependence on~$t$ is then restricted to the incident sensitivity component.

\clearpage
\section{Additional simulation results}
\label{app:sims}

\subsection{Simulations with different settings}
This section presents additional simulations assessing the NPB framework. We use different event time distributions, sample sizes and censoring rates.

\begin{figure}[h!]
	\centering
	\includegraphics[width=0.85\linewidth]{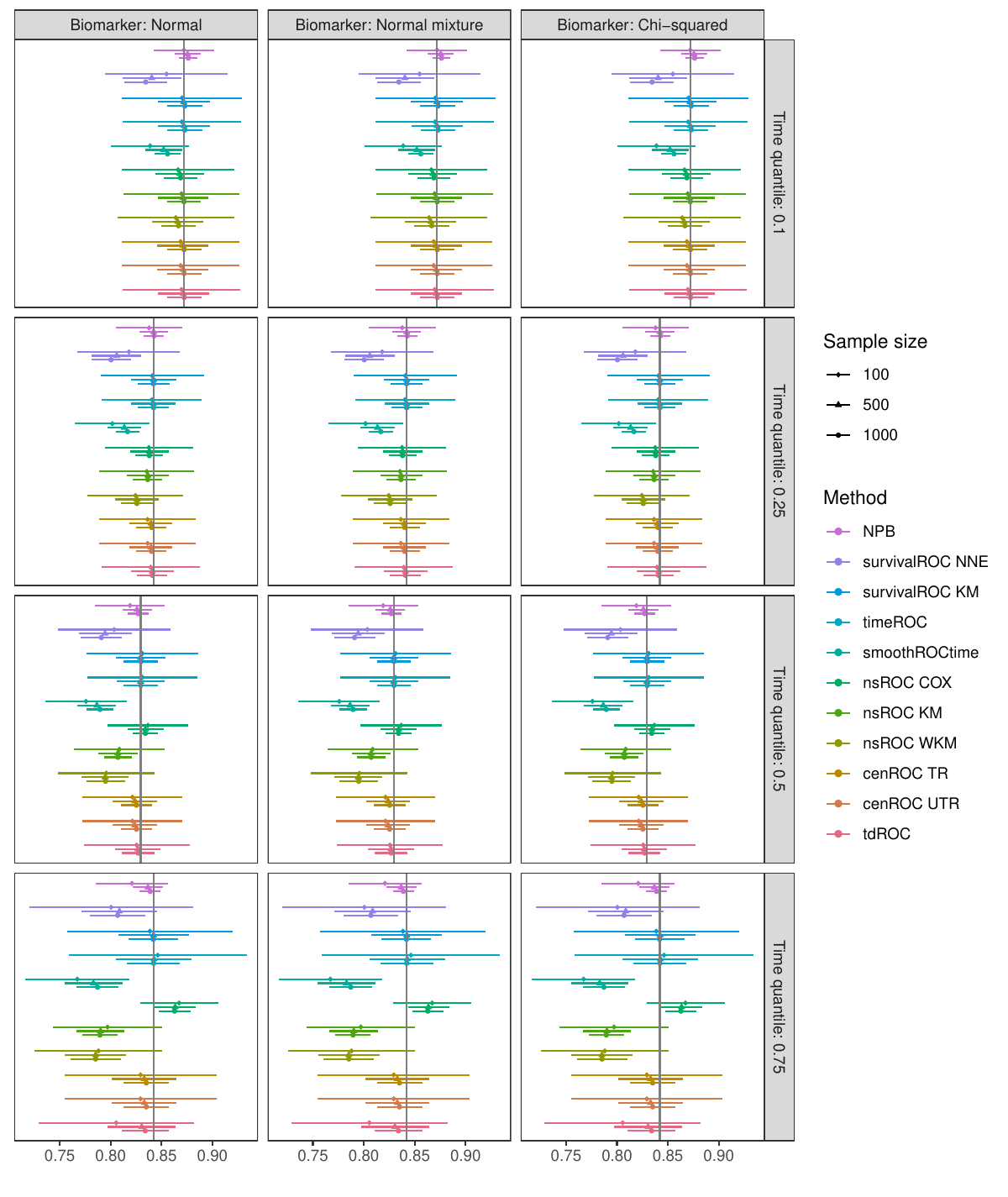} 
	\caption{Mean and standard deviation of the unconditional AUC for each method with sample sizes of $N = \{100, 500, 1000\}$, Lognormal event time distribution, 50\% censoring rate and a correlation of $\rho = -0.7$ between transformed biomarker and event time distributions. The evaluation is across different biomarker distributions (Normal, Normal mixture, Chi-squared) and time quantiles (0.1, 0.25, 0.5, 0.75). }
	\label{fig:app_sim_auct}
\end{figure}

\begin{figure}[h!]
	\centering
	\includegraphics[width=0.85\linewidth]{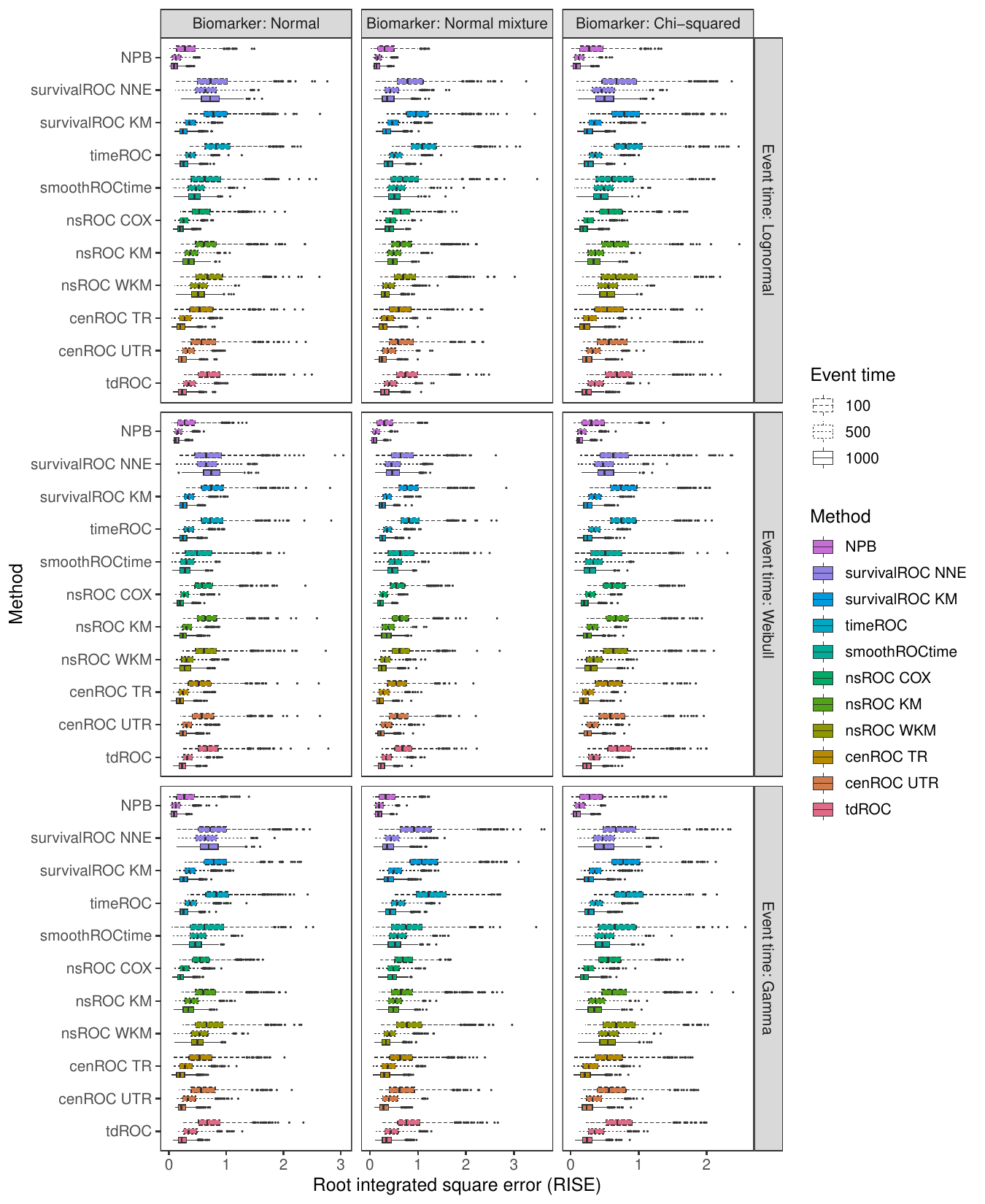} 
	\caption{Distribution of root integrated squared errors (RISE) for unconditional ROC estimators at the median time quantile with sample sizes of $N \in \{100, 500, 1000\}$, event time distributions of $\{\text{Lognormal}, \text{Weibull}, \text{Gamma}\}$, 50\% censoring rate and a correlation of $\rho = -0.7$ between transformed biomarker and event time distributions.}
	\label{fig:app_sim_roct}
\end{figure}

\begin{figure}[h!]
	\centering
	\includegraphics[width=0.85\linewidth]{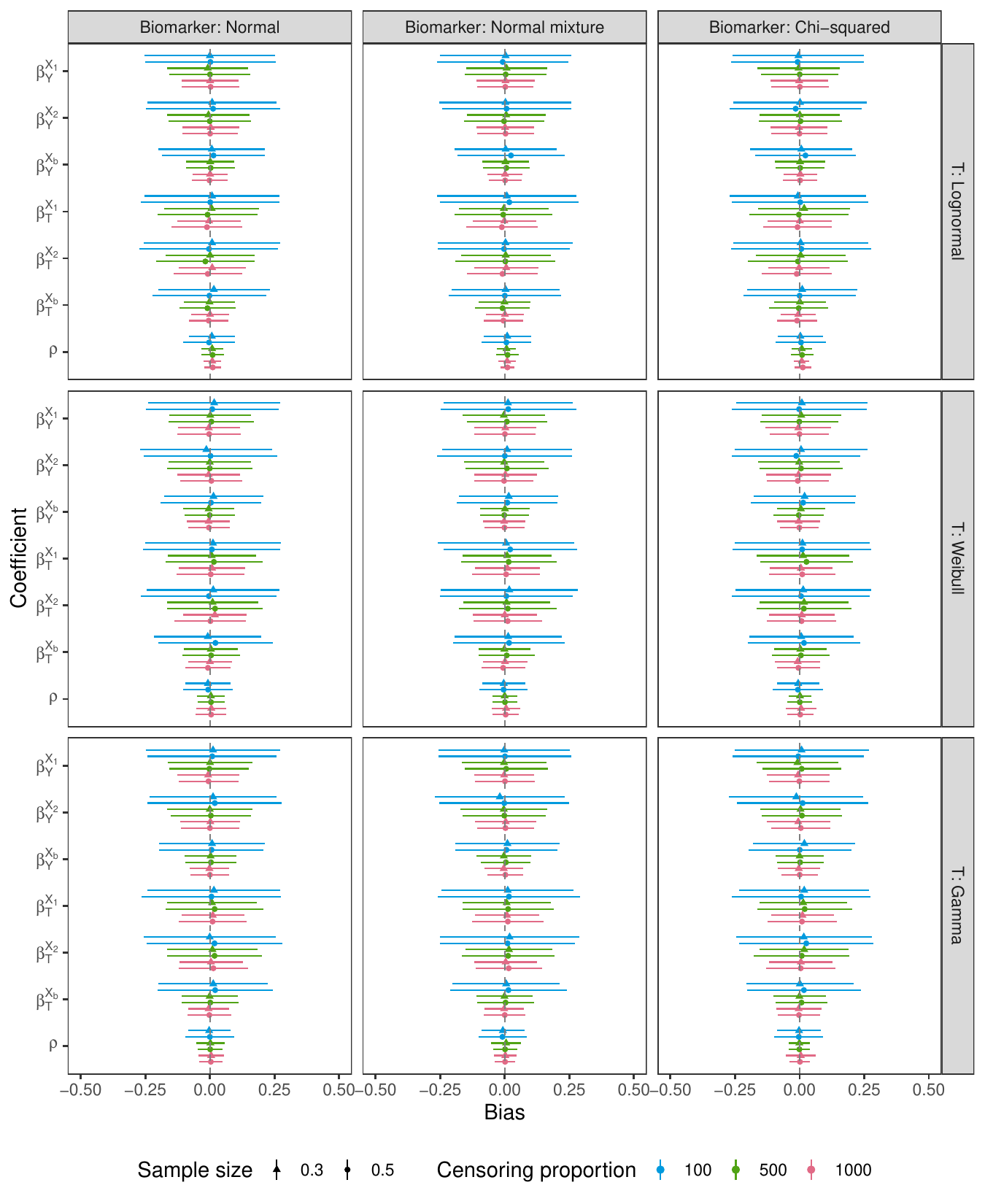} 
	\caption{Mean and standard deviation of bias for coefficients for the NPB method with sample sizes of $N \in \{100, 500, 1000\}$, censoring rates of 30\% and 50\%, and a correlation of $\rho = -0.5$. Results are presented for varying biomarker distributions \{Normal, Normal mixture, Chi-squared\} and event time distributions \{Lognormal, Weibull, Gamma\}. }
	\label{fig:sim_auctx}
\end{figure}

\clearpage
\subsection{Simulations with dependence misspecification}
\label{sec:misspec}

We investigated how the NPB model performs when there is a misspecification of the dependence structure. For this, we considered an alternative DGP proposed by \AYcite{Rodr\'{\i}guez-\'Alvarez et al.}{rodriguez2016nonparametric}. In this setting the true relationship between the biomarker and event time deviates from the assumed copula-based dependence structure.

The covariate $\rX$ is generated from a normal distribution with mean $1$ and variance
$1$, i.e., $\rX \sim N(1, 1)$. The biomarker $Y$ is then generated from a conditional normal distribution with mean $X = x$ and variance 1, i.e., $Y\sim N(x, 1)$. Finally, the survival time $T$ is generated from a proportional hazards Weibull model, conditional on the biomarker and covariate. This corresponds to an accelerated failure time (AFT) model
\begin{align*}
	\log(T) = y + 0.5 x + Z,
\end{align*}
where $Z$ follows a minimum extreme value distribution. 

\begin{figure}[h!]
	\centering
	\includegraphics[width=\linewidth]{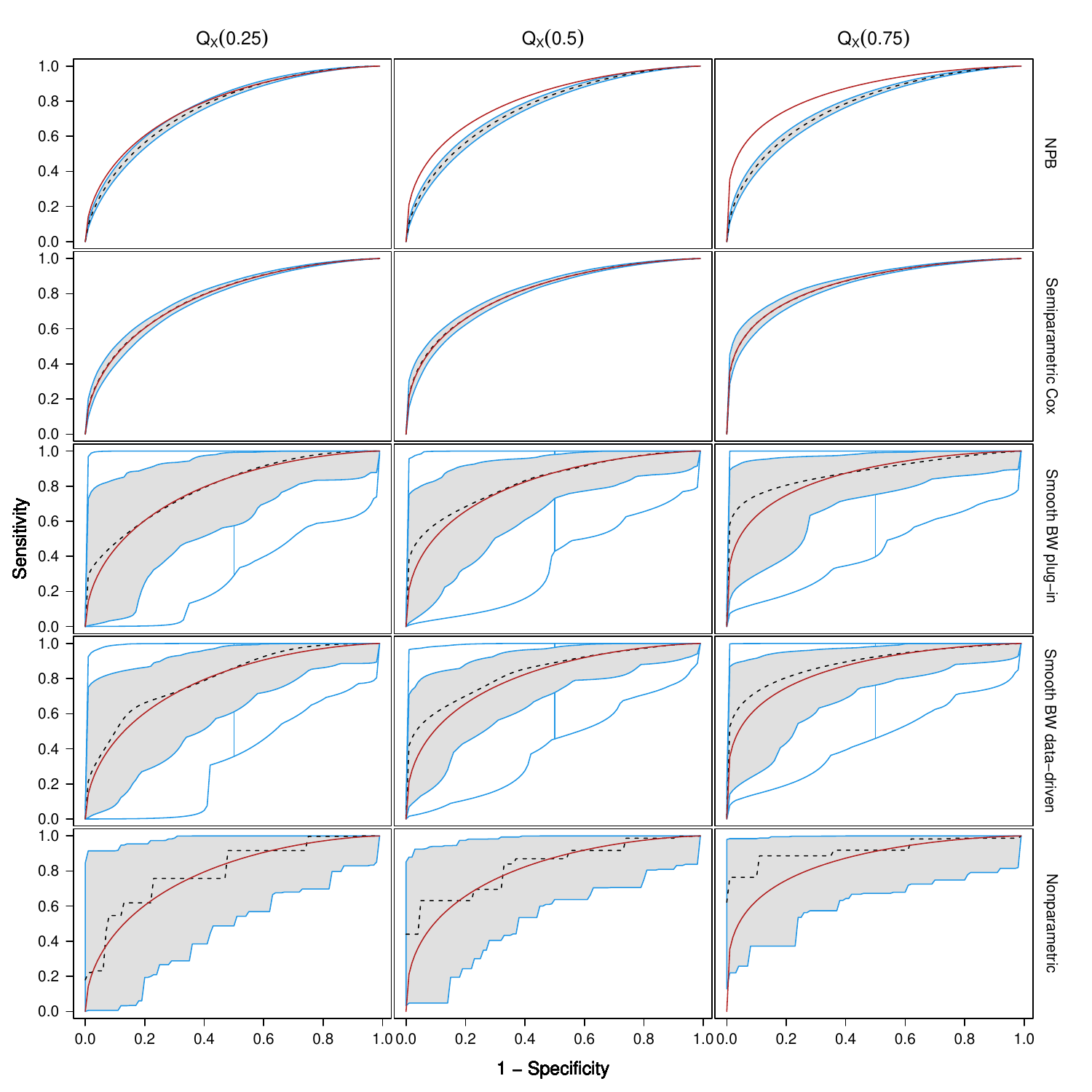} 
	\caption{Functional boxplots for covariate-specific time dependent ROC curves at the median time quantile, with a sample size of $N = 500$, Weibull event time distribution,
	$30\%$ censoring rate, and a correlation of $\rho = -0.5$ between transformed biomarker and event time distributions. The red line indicates the true covariate-specific ROC curve for each case.}
	\label{fig:sim_roctx_ms}
\end{figure}

To assess performance, we generated $1000$ replications, with each dataset
consisting of $500$ observations with a censoring rate of $30\%$.  We
evaluated the covariate-specific cumulative-dynamic time-dependent ROC
curves $\ROCcd_t(x)$ $25$th, $50$th, and $75$th percentiles of the covariate. 
Figure~\ref{fig:sim_roctx_ms} displays functional boxplots of the results
and compares the different covariate-specific time-dependent ROC curve
methods for a single time point of $1.5$.

The NPB model shows low bias for low quantiles of $X$. However, the bias increases across higher quantiles. The proportional hazards-based semiparametric method of \AYcite{Song and Zhou}{song2008semiparametric} outperformed other methods, since it correctly aligns with the true DGP. The nonparametric kernel-based approach was largely unbiased across quantiles but exhibited higher variability.

These results suggest that while the NPB model effectively captures correlation-driven dependencies between biomarkers and event times, it struggles when the dependence follows a direct functional form. That is, when $T$ is assumed to be a function of $Y$ and $X$. This highlights a fundamental bias-variance trade-off for method selection. Model-based approaches tend to provide low-variance, unbiased estimates when correctly specified but can introduce bias when the true DGP differs from the assumed model. In contrast, fully nonparametric methods are asymptotically unbiased but often suffer from high variability in finite samples, making estimation less stable. Additionally, nonparametric approaches have problems in incorporating multiple covariates. Ultimately, the choice of method depends on the specific application and the needs of the target audience. 

\clearpage
\section{Additional application results}

\begin{figure}[h!]
	\centering
	\includegraphics[width=0.85\linewidth]{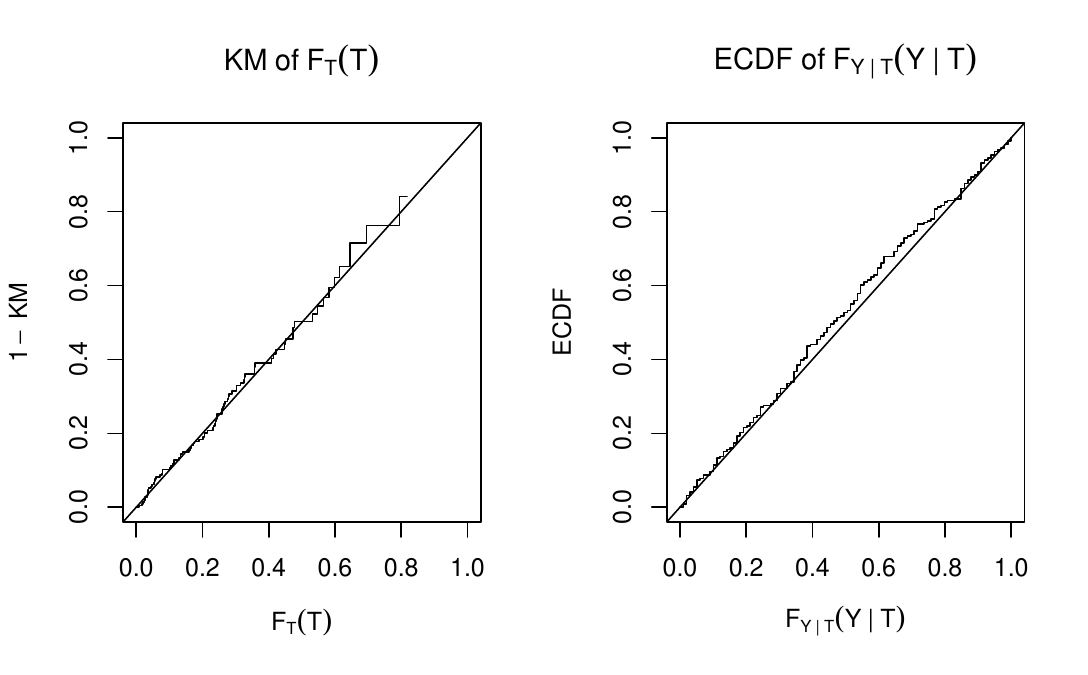} 
	\caption{QQ-PIT diagnostic plots for assessing model calibration without covariates. Left: Kaplan–Meier estimate of PIT values for survival time. Right: ECDF of conditional PIT values for the biomarker, accounting for right censoring.}
	\label{fig:pitu}
\end{figure}

\begin{figure}[h!]
	\centering
	\includegraphics[width=0.85\linewidth]{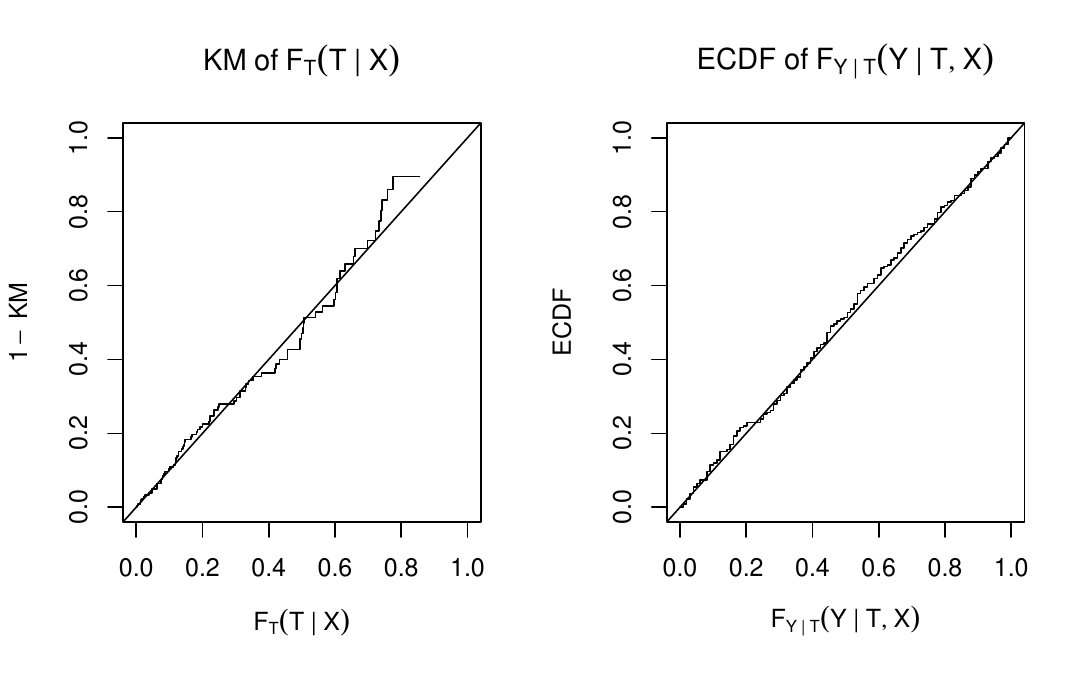} 
	\caption{QQ-PIT diagnostic plots for assessing model calibration with covariates. Left: Kaplan–Meier estimate of PIT values for survival time given covariates. Right: ECDF of conditional PIT values for the biomarker, adjusted for covariates and right censoring.}
	\label{fig:pitc}
\end{figure}

\begin{figure}[h!]
	\centering
	\includegraphics[width=0.9\linewidth]{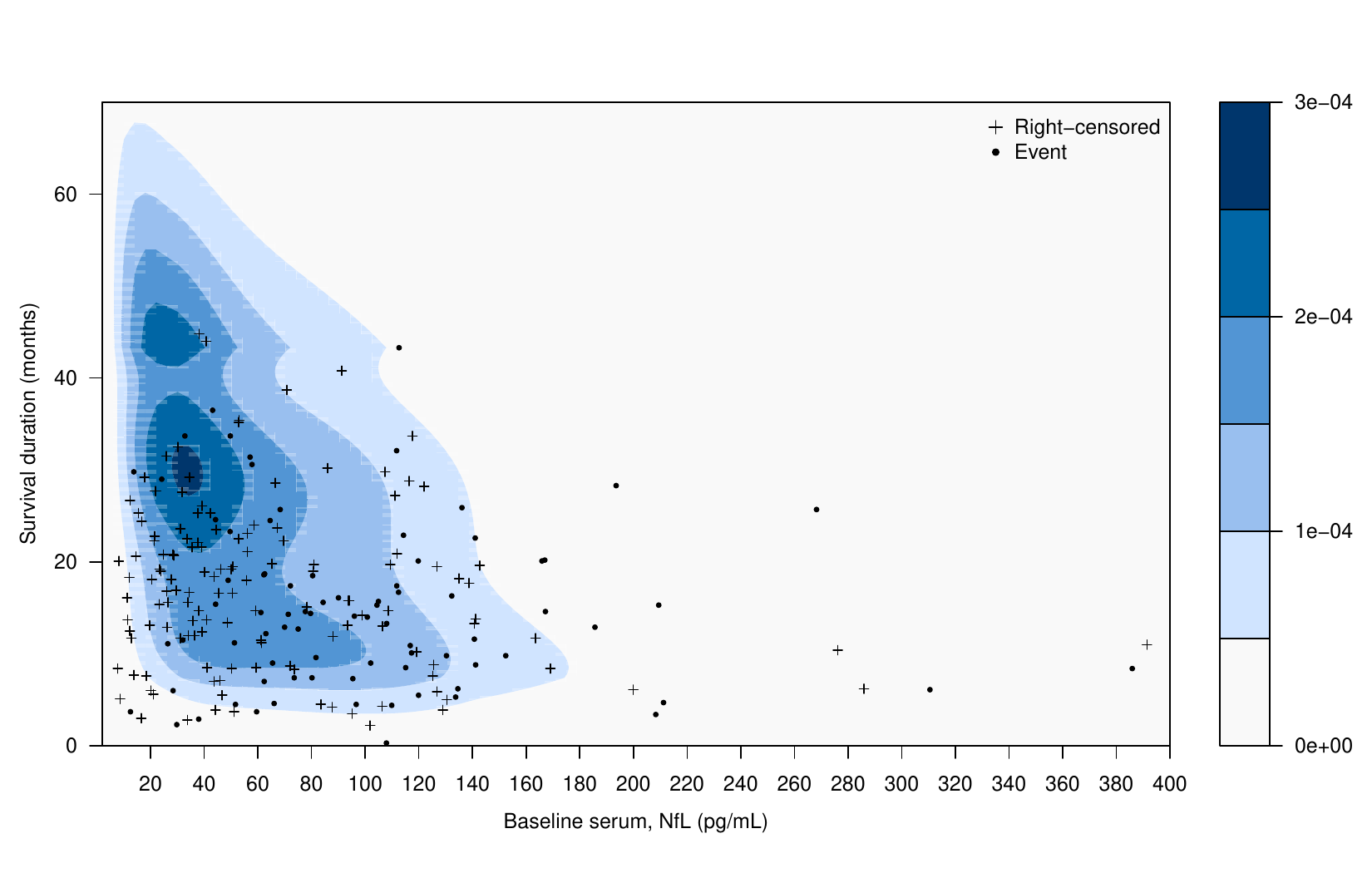} 
	\caption{Estimated bivariate density of baseline serum neurofilament light~(NfL) concentration (pg/mL) and survival time (in months since baseline) overlayed with observed data. Right censored subjects are indicated with~``+'' and subjects who reached the endpoint are marked by~``$\sbullet[0.5]$''.}
	\label{fig:densu}
\end{figure}

\begin{figure}[h!]
	\centering
	\includegraphics[width=0.85\linewidth]{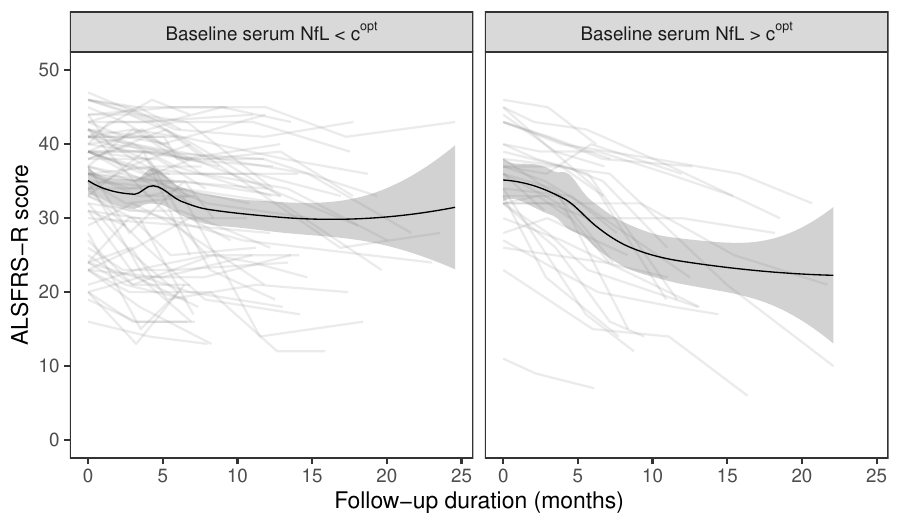} 
	\caption{Spaghetti plot of ALSFRS-R scores over follow-up stratified by the optimal baseline serum neurofilament light concentration threshold for predicting 12-month survival.}
	\label{fig:longit}
\end{figure}

\begin{figure}[h!]
	\centering
	\includegraphics[width=\linewidth]{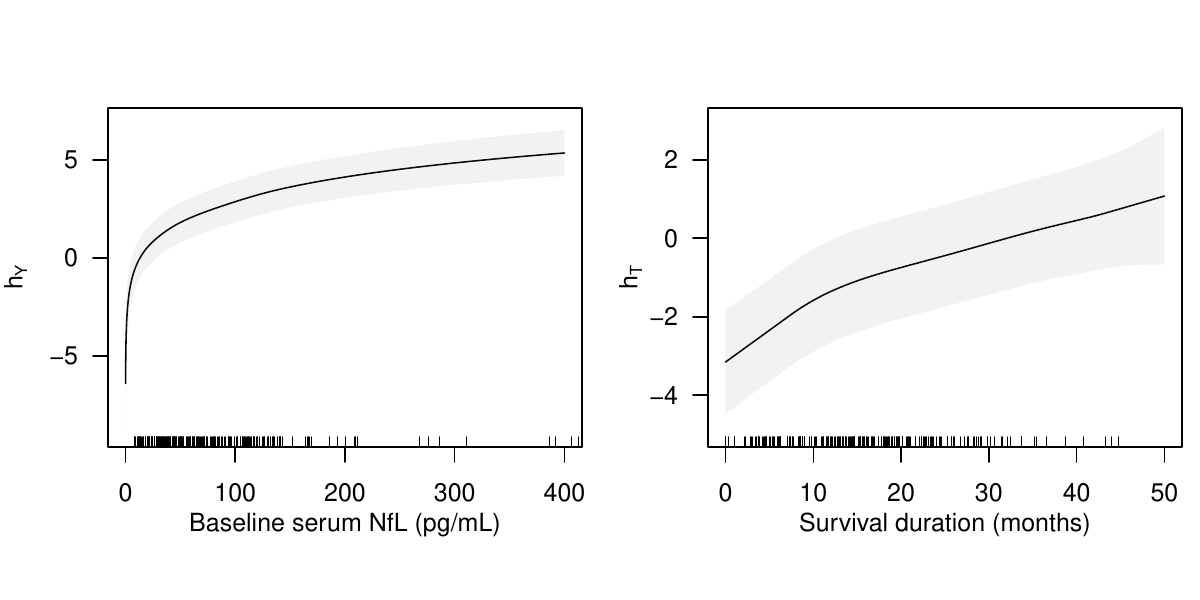} 
	\caption{Estimated baseline transformation functions for baseline serum neurofilament light~(NfL) concentration~$h_Y(y \mid \hat{\parm}_Y)$ and survival time $h_T(t \mid \hat{\parm}_T)$.}
	\label{fig:trafo}
\end{figure}

 \end{appendix}

\clearpage

\end{document}